\documentclass[utf8]{frontiersSCNS} 

\usepackage{url,hyperref,lineno,microtype,subcaption}
\usepackage[onehalfspacing]{setspace}

\usepackage{pdflscape}
\usepackage{mathrsfs}

\def\keyFont{\fontsize{8}{11}\helveticabold }
\def\firstAuthorLast{D. M. Bowman} 
\def\Authors{Dominic M. Bowman\,$^{1,*}$}

\def\Address{$^{1}$Institute of Astronomy, KU Leuven, Celestijnenlaan 200D, B-3001 Leuven, Belgium}
\def\corrAuthor{Dominic M. Bowman}

\def\corrEmail{dominic.bowman@kuleuven.be}

\begin{document}
\onecolumn
\firstpage{1}

\title[Asteroseismology of high-mass stars]{Asteroseismology of high-mass stars: \\ new insights of stellar interiors with space telescopes} 

\author[\firstAuthorLast ]{\Authors} 
\address{} 
\correspondance{} 

\extraAuth{}

\maketitle


\begin{abstract}
Massive stars are important metal factories in the Universe. They have short and energetic lives, and many of them inevitably explode as a supernova and become a neutron star or black hole. In turn, the formation, evolution and explosive deaths of massive stars impact the surrounding interstellar medium and shape the evolution of their host galaxies. Yet the chemical and dynamical evolution of a massive star, including the chemical yield of the ultimate supernova and the remnant mass of the compact object, strongly depend on the interior physics of the progenitor star. We currently lack empirically calibrated prescriptions for various physical processes at work within massive stars, but this is now being remedied by asteroseismology. The study of stellar structure and evolution using stellar oscillations -- asteroseismology -- has undergone a revolution in the last two decades thanks to high-precision time series photometry from space telescopes. In particular, the long-term light curves provided by the MOST, CoRoT, BRITE, Kepler/K2 and TESS missions provided invaluable data sets in terms of photometric precision, duration and frequency resolution to successfully apply asteroseismology to massive stars and probe their interior physics. The observation and subsequent modelling of stellar pulsations in massive stars has revealed key missing ingredients in stellar structure and evolution models of these stars. Thus asteroseismology has opened a new window into calibrating stellar physics within a highly degenerate part of the Hertzsprung--Russell diagram. In this review, I provide a historical overview of the progress made using ground-based and early space missions, and discuss more recent advances and breakthroughs in our understanding of massive star interiors by means of asteroseismology with modern space telescopes.

\tiny
 \keyFont{ \section{Keywords:} asteroseismology, stars: interiors, stars: oscillations, stars: evolution, stars: rotation, stars: massive, stars: early-type} 
\end{abstract}


\section{Introduction}
\label{section: introduction}

Stars are the essential building blocks of planetary systems, stellar clusters and galaxies. The lives and energetic deaths of massive stars -- i.e. those with birth masses larger than approximately eight times that of the Sun (8 M$_{\odot}$) -- play a pivotal role in shaping the Universe \citep{Maeder2000a, Maeder_rotation_BOOK, Langer2012, Kippenhahn_BOOK}. Massive stars were amongst the first stars in our Universe \citep{Bromm2004c, Bromm2009b}, and are progenitors of core-collapse supernovae and gamma-ray bursts \citep{Heger2003e, Smartt2009b, Tanvir2009, Modjaz2019a}. The properties of massive stars allow them to be observed at large distances (see e.g. \citealt{Stark2016}), hence allow us to study the early epochs of the Universe including the re-ionisation of the Universe and the formation of the first galaxies \citep{Bromm2004c, Robertson_B_2010d}.

Massive stars typically form in dense, cold and large molecular clouds with one of the important signatures of massive star formation being giant filament structures and powerful bi-polar outflows when they are embedded in such dense clouds --- see the recent review by \citet{Rosen2020a}. After the formation phase, massive stars enter the so-called main sequence phase of stellar evolution, which is defined by the onset of hydrogen fusion in the core via the CNO cycle whilst maintaining hydrostatic equilibrium. The length of the main sequence is governed by the nuclear time scale, with more massive stars having shorter main sequence life times. During their lives, massive stars produce intense radiation fields from their high luminosities and experience line-driven winds, which together play major roles in the shaping of their environment \citep{Langer2012, Kippenhahn_BOOK}. The majority of massive stars experience an explosive death as a supernova, which provides mechanical and chemical feedback to the interstellar environment \citep{MacLow2004b, deRossi2010d, Hopkins_P_2014a, Stark2016, Crowther2016} and can trigger a new generation of stars and planets. The exotic compact remnants of massive star evolution are neutron stars and black holes, with the latter being end products for many of the higher-mass progenitor stars. These remnants facilitate important tests of Einstein’s theory of General Relativity and the study of the Universe using gravitational waves when they coalesce \citep{Abbott_B_2016d, Abbott_B_2019m}. Hence, understanding the evolution of massive stars and their roles as supernovae and gravitational-wave progenitors represent fundamental questions in astronomy \citep{Smartt2009b, Langer2012}. It is especially important to understand massive star evolution since there is a strong dependence of a supernova's chemical yield and the mass of the remnant on the interior physics of the progenitor star \citep{Nomoto2006, Langer2012, Stark2016}, and because of the large diversity in observed supernovae light curves \citep{Dessart2019b}.

Despite the importance of massive stars in our Universe, their physics is not yet fully understood. There are still many questions spanning all evolutionary phases, and specifically how their formation, evolution and inevitable deaths differ to those of the more common intermediate- and low-mass stars \citep{Langer2012}. A major shortcoming of current stellar evolution models is that they contain large theoretical uncertainties for massive stars, which is evident already during the earliest phases of stellar evolution including the main sequence. Consequently these uncertainties propagate and strongly impact the post-main sequence stage of stellar evolution \citep{Maeder2000a, Ekstrom2012a, Chieffi2013}. Hence the power of models for predicting if a massive star will explode as a supernova, the corresponding chemical yield and the mass of the compact remnant are limited by the accuracy of models reproducing the observed properties of stars prior to them exploding as supernovae.

Since massive stars have convective cores and radiative envelopes during the main sequence, the physics and numerical implementation of convection and convective-boundary mixing is crucial in determining their core masses and subsequent evolution \citep{Gabriel2014, Georgy2014a, Paxton2018, Paxton2019}. The mixing profile at the interface of convective and radiative regions, and the mixing profile within the envelope directly impact the amount of hydrogen available for nuclear burning. With more internal mixing, a massive star experiences a longer main sequence and produces a larger helium core mass at the end of the main sequence since fresh hydrogen from the envelope is readily supplied to the convective core \citep{Miglio2008a, Kippenhahn_BOOK, Pedersen2018a, Michielsen2019a}. In the case of single, slowly rotating and non-magnetic massive stars, it is the internal mixing profile and helium core mass that dictates the evolution beyond the main sequence and determines the ultimate end state. However, probing the physical processes beneath the opaque surfaces of massive stars is practically impossible using standard methods and techniques in astronomy, such as spectroscopy.

Rotation plays a major role among massive stars, specifically because rotationally induced mixing is expected within their interiors \citep{Zahn1992, Maeder2000a}. Yet prescriptions for such mixing profiles are currently assumed in models and controlled by numerous free parameters that have yet to be empirically calibrated. This represents a large source of uncertainty within theoretical evolution models when estimating the masses and ages of high-mass stars (see e.g. \citealt{Aerts2020a**} and \citealt{Serenelli2020a*} for recent detailed reviews). The combined influence of various rotation rates, different metallicity regimes and mass loss through stellar winds also introduce strong degeneracies within evolutionary models of massive stars \citep{Maeder2000a, Georgy2011a, Ekstrom2012a, Georgy2013c, Chieffi2013, Groh2019a}. Furthermore, based on dedicated studies such as MiMeS, \citep{Wade2016a}, the BOB campaign \citep{Morel2015b}, and the BRITE spectropolarimetric survey \citep{Neiner2017b}, approximately 10\,\% of massive stars are inferred to host a large-scale magnetic field with polar field strengths that range from approximately 100~G up to a few kG. The degeneracies within evolutionary models are also more complex when dealing with the presence of magnetic fields in massive stars \citep{Alecian_E_2014a, Shultz2018d, Keszthelyi2019, Keszthelyi2020a}. 

It is known that many massive stars are members of multiple systems, which interact over the course of their lifetimes, so theoretical uncertainties are further compounded by the effects of binarity and mass transfer \citep{Podsiadlowski1992c, Sana2012b, deMink2013, Duchene2013b, Moe2017a}. However, despite the complexities associated with rotation, metallicity, mass loss and magnetic fields, massive stars in multiple systems have proven extremely useful in probing and mitigating model parameter uncertainties \citep{Guinan2000, Hilditch2005b, Torres2010a, Tkachenko2014a, Tkachenko2016, Almeida2015, Abdul-Masih2019a, Johnston2019b, Mahy2020a, Mahy2020b}. This is primarily because binary studies have the potential to provide masses and radii from the stars' relative orbital motion around a common centre of mass. Moreover, accurate, absolute and model-independent masses and radii are achievable in the case of eclipsing binary systems (see e.g. \citealt{Tkachenko2020a} and \citealt{Southworth2020a}) owing to the ability to simultaneously model spectroscopic radial velocities together with the eclipse depths in the light curve of a binary system. 

To truly maximise the predictive power of evolutionary models for massive stars, it is essential to calibrate their physical prescriptions and parameters using stringent observational constraints on stellar interiors. One of the most successful and novel methodologies for this is called asteroseismology, which uses the resonant oscillation frequencies of stars to probe their structure --- see the research monograph by \citet{ASTERO_BOOK}. Until recently, most observational studies of massive stars have focussed on the determination of global and/or average properties of these stars, such as the effective temperature and surface gravity derived from spectroscopy being used to estimate masses and ages. On the other hand, the recent space photometry revolution has truly brought asteroseismology to the forefront of astronomy as a means to calibrate stellar evolution theory across the Hertzsprung--Russell (HR) diagram \citep{Chaplin2013c, Hekker2017a, Garcia_R_2019, Aerts2020a**}.

In this review, I discuss the progress that has been made in constraining the interiors of massive stars by means of asteroseismology and the space photometry revolution made possible thanks to modern space missions. In section~\ref{section: asteroseismology}, I provide a brief overview of asteroseismology, its methodology, and the types of pulsating massive stars it can be applied to. In section~\ref{section: space photometry}, an overview of the space telescopes that have led to the space photometry revolution is described. In section~\ref{section: constraints}, I discuss the recent advances in massive star interiors made by means of asteroseismology, and section~\ref{section: waves} describes the state-of-the-art of variability studies in some of the most massive stars. I finish by discussing the current challenges and future prospects for asteroseismology of high-mass stars in section~\ref{section: conclusions}.


\section{Asteroseismology}
\label{section: asteroseismology}

A powerful method for probing and constraining the physics of stellar interiors is asteroseismology, which uses stellar oscillations to probe the physics of stellar structure \citep{ASTERO_BOOK}. The pulsation modes of stars are standing waves exhibiting nodes and anti-nodes and are described by spherical harmonics. In the case of non-rotating and non-magnetic stars, the wavefunctions of stellar pulsations are separable into the radial and angular directions. The radial parts of the wavefunction solutions are characterised by the radial order $n$. Whereas the angular dependence is characterised by the angular degree $\ell$ (number of surface nodes), and the azimuthal order $m$ (where $|m|$ is the number of surface nodes that are lines of longitude). The simplest example of a pulsation mode is a radial mode for which $\{\ell, m\} = 0$ such that the surface of a star expands and contracts during a pulsation cycle. More complex examples of pulsation modes include non-radial modes (i.e. $\ell > 0$), for which the indices $\ell$ and $m$ define the surface geometry of the oscillation. As an example, the axisymmetric dipole mode (i.e. $\{\ell, m\} = \{1, 0\}$) has the stellar equator as a node. Thus, the northern and southern hemispheres of a star expand and contract in anti-phase with one another.

Although they have a common structure comprising a convective core and a radiative envelope during the main sequence, there are different types of pulsations that can be excited in massive stars. In general, however, the excitation mechanism of pulsation modes has been shown to be the heat-engine mechanism operating in the local maximum of the Rosseland mean opacity caused by iron-group elements --- the so-called Z-bump \citep{Dziembowski1993e, Dziembowski1993f, Gautschy1993a, Pamyat1999b, Miglio2007a}. This $\kappa$-mechanism gives rise to pulsation modes with properties and excitation physics which depend on the host star's mass, age and chemical composition. There are two main types of pulsation modes excited by the $\kappa$~mechanism in massive stars, which are defined based on their respective restoring force: pressure (p) modes and gravity (g) modes.

\vspace{0.5cm}


\subsection{Pressure modes}
\label{subsection: pressure modes}

Pressure (p) modes are standing waves for which the pressure force acts as a restoring force \citep{ASTERO_BOOK}. Typically, p~modes have high frequencies (i.e. pulsation periods of order several hours in massive stars), can be radial or non-radial and are mostly sensitive to the radiative envelopes of massive stars. For radial p~modes, the entire interior of the star acts as a pulsation cavity, with the centre of a star being a node and its surface an anti-node. In the cases of non-radial p~modes, the depth of the pulsation cavity from the surface is determined by the local adiabatic sound speed $c(r)$. A non-radial pulsation mode encounters an increasing $c(r)$ when travelling inward from the stellar surface, which causes it to travel faster and be refracted. The depth a non-radial p~mode can reach is called its turning radius, $r_{t}$, which is proportional to $\sqrt{\ell(\ell + 1)}$ and defined outwards from the centre of the star \citep{ASTERO_BOOK}. Thus, higher degree p~modes have smaller pulsation cavities that are more sensitive to the stellar surface. The power of asteroseismology is that each pulsation has a cavity defined by a star's structure, such that each pulsation mode can be used as a direct probe of the physical processes at work within its cavity.

If the radial orders of p~modes are sufficiently large such that the modes satisfy $n \gg \ell$, which is called the asymptotic regime, the pulsations are approximately equally-spaced in frequency \citep{Tassoul1980}. Deviations from a constant frequency spacing are possible, and become more prevalent for more evolved stars. During the main sequence phase of evolution the radius of a massive star increases and the core contracts, which drives the g- and p-mode pulsation cavities closer to one another as a result of an increasing Brunt-V{\"a}is{\"a}l{\"a} frequency \citep{ASTERO_BOOK}. Consequently, this can cause a form of pulsation mode interaction called avoided crossings, in which p and g modes can exchange character whilst retaining their identities if their frequencies approach one another \citep{Osaki1975}. In more evolved cases, such as post main-sequence stars, the evanescent region between the p- and g-mode cavities inside massive stars decreases. This can allow p and g~modes to couple with each other and form mixed modes, which are modes with the character of a p~mode in the envelope and the character of a g~mode in the deep interior \citep{ASTERO_BOOK}. The regularities of asymptotic p~modes in the amplitude spectra of low- and intermediate-mass stars has greatly simplified the issue of mode identification and facilitated asteroseismology for low-mass stars (see e.g. \citealt{Chaplin2013c, Hekker2017a, Garcia_R_2019}), but are rarely observed in massive stars (see e.g. \citealt{Belkacem2010a, Degroote2010b}). Such high-radial order p~modes are generally not expected for massive stars owing to the excitation physics of the $\kappa$-mechanism being inefficient in driving such modes in massive stars \citep{Dziembowski1993e, Dziembowski1993f, Gautschy1993a, Pamyat1999b, Miglio2007a}.

\begin{figure}[ht!]
\centering
\includegraphics[width=0.5\textwidth,clip]{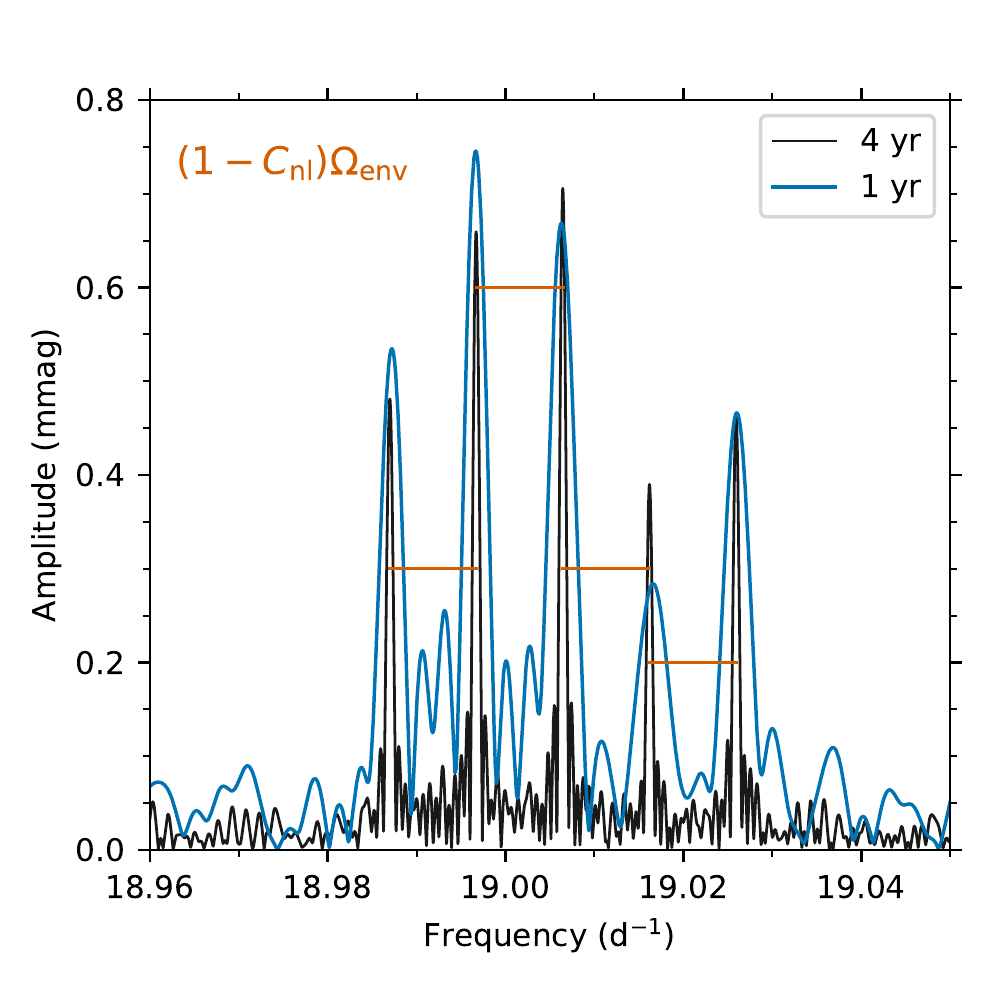}      
\caption{Example of rotational splitting of quadrupole p~modes into a quintuplet using both 1-yr and 4-yr light curves of the star KIC~11145123 \citep{Kurtz2014}. Horizontal red lines correspond to the rotational splitting value of the modes.}
\label{figure: rotation}
\end{figure}

In the presence of rotation the frequency degeneracy of non-radial pulsation modes with respect to $m$ is lifted, which serves as a unique method of mode identification in certain pulsating stars. The simplest case is for stars that rotate (very) slowly and rigidly -- i.e. with a uniform interior rotation angular frequency $\Omega$ -- such that the splitting of non-radial pulsation frequency, $\omega_{n \ell m}$, is given by

\begin{equation}
\omega_{n \ell m} = \omega_{n \ell} + m\left(1 - C_{n \ell}\right) \Omega ~ , 
\label{equation: rotation}
\end{equation}

\noindent where $C_{n \ell}$ is the Ledoux constant which sets the size of the splitting due to the Coriolis force. In this idealised example, the result of Equation~\ref{equation: rotation} produces a multiplet of pulsation frequencies separated by the stellar rotation frequency in the amplitude spectrum for p~modes of high radial order or high-angular degree since $C_{n\ell} \simeq 0$ in such cases \citep{ASTERO_BOOK}. An example of rotationally-split quadrupole p~modes is shown in Fig.~\ref{figure: rotation}, using the example of KIC~11145123 originally discovered by \citet{Kurtz2014}. The amplitude spectrum of the resultant quintuplet split by rotation shown in Fig.~\ref{figure: rotation} uses both 1-yr and 4-yr light curves to emphasise the significant improvement in the resolving power of longer light curves for asteroseismic studies of rotation. Therefore, if the rotation rate is sufficiently slow, p-mode multiplets serve as a means of determining the interior rotation rates of stellar envelopes using an almost model-independent methodology.

Beyond the first-order perturbative approach for including the Coriolis force in slow and rigid rotators given in Equation~\ref{equation: rotation}, second- and third-order perturbative formalisms have been discussed by, for example, \citet{Dziembowski1992c}, \citet{Daszy2002b} and \citet{Suarez2010a}. As described by \citet{Suarez2010a}, it is important to note that the first-order perturbative treatment of the Coriolis force applied to p~modes is only applicable for stars with rotation velocities below approximately 15\,\% of their critical breakup velocity, with faster rotating stars requiring more complex formalisms. 

\vspace{0.5cm}


\subsection{Gravity modes}
\label{subsection: gravity modes}

Gravity (g) modes are standing waves for which buoyancy (i.e. gravity) acts as a restoring force \citep{ASTERO_BOOK}. Typically, g~modes have low frequencies, can only be non-radial and are mostly sensitive to the deep interiors of massive stars near their convective cores. In the asymptotic regime, g~modes are equally spaced in period \citep{Tassoul1980}, and exhibit a characteristic period $\Pi_{0}$. In the case of a non-rotating and chemically-homogeneous star, $\Pi_{0}$ can be calculated from the individual g-mode periods, $P_{n, \ell}$, given by

\begin{equation}
P_{n \ell} = \frac{\Pi_{0}}{\sqrt{\ell\left(\ell+1\right)}} \left( |n| + \alpha \right) ~ ,
\label{equation: periods}
\end{equation}

in which $\alpha$ is a phase term independent of the mode degree, $\ell$, and

\begin{equation}
\Pi_{0} = 2\pi^{2} \left( \int_{r_1}^{r_2} N(r) \frac{\rm dr}{r} \right)^{-1} ~ ,
\label{equation: pi}
\end{equation}

\noindent where $r_{1}$ and $r_{2}$ are the inner and outer boundaries of the g-mode pulsation cavity, and $N(r)$ is its Brunt-V{\"a}is{\"a}l{\"a} frequency. Thus, Equation~\ref{equation: periods} defines a constant spacing in period for g~modes of the same angular degree, $\ell$, and consecutive radial order, $n$. Equation~\ref{equation: pi} demonstrates that the characteristic period, $\Pi_{0}$, is largely determined by the Brunt-V{\"a}is{\"a}l{\"a} frequency, $N(r)$, which has a strong dependence on the mass of the convective core, and hence the mass and age of a star \citep{Miglio2008b}.

\vspace{0.5cm}


\subsubsection{Interior rotation}
\label{subsubsection: g-mode rotation}

Since all massive stars rotate to some extent, the Coriolis force is also a dominant restoring force for gravity modes. Therefore, it is more appropriate to describe massive stars having gravito-inertial modes, for which both the Coriolis force and buoyancy are important. This is particularly true for pulsation modes with frequencies in the co-rotating frame below twice the rotation frequency (see \citealt{Aerts2019b}). As discussed in detail by \citet{Bouabid2013}, the period spacing increases with period in the co-rotating frame for prograde modes, and decreases in the inertial frame. This is because in the co-rotating frame the effective $\ell(\ell+1)$ (cf. Equation~\ref{equation: periods}) for prograde sectoral modes decreases with rotation due to the effect of Coriolis force. Whereas in the inertial frame an increasing period spacing is caused by the frequency increase (i.e. period decrease) due to the effect of advection $|m|\Omega$. Therefore, for rotating stars as viewed in the inertial frame by an observer, one expects a decreasing period spacing for prograde g~modes and an increasing period spacing for retrograde g~modes. 

\begin{figure}[ht!]
\centering
\includegraphics[width=0.95\textwidth,clip]{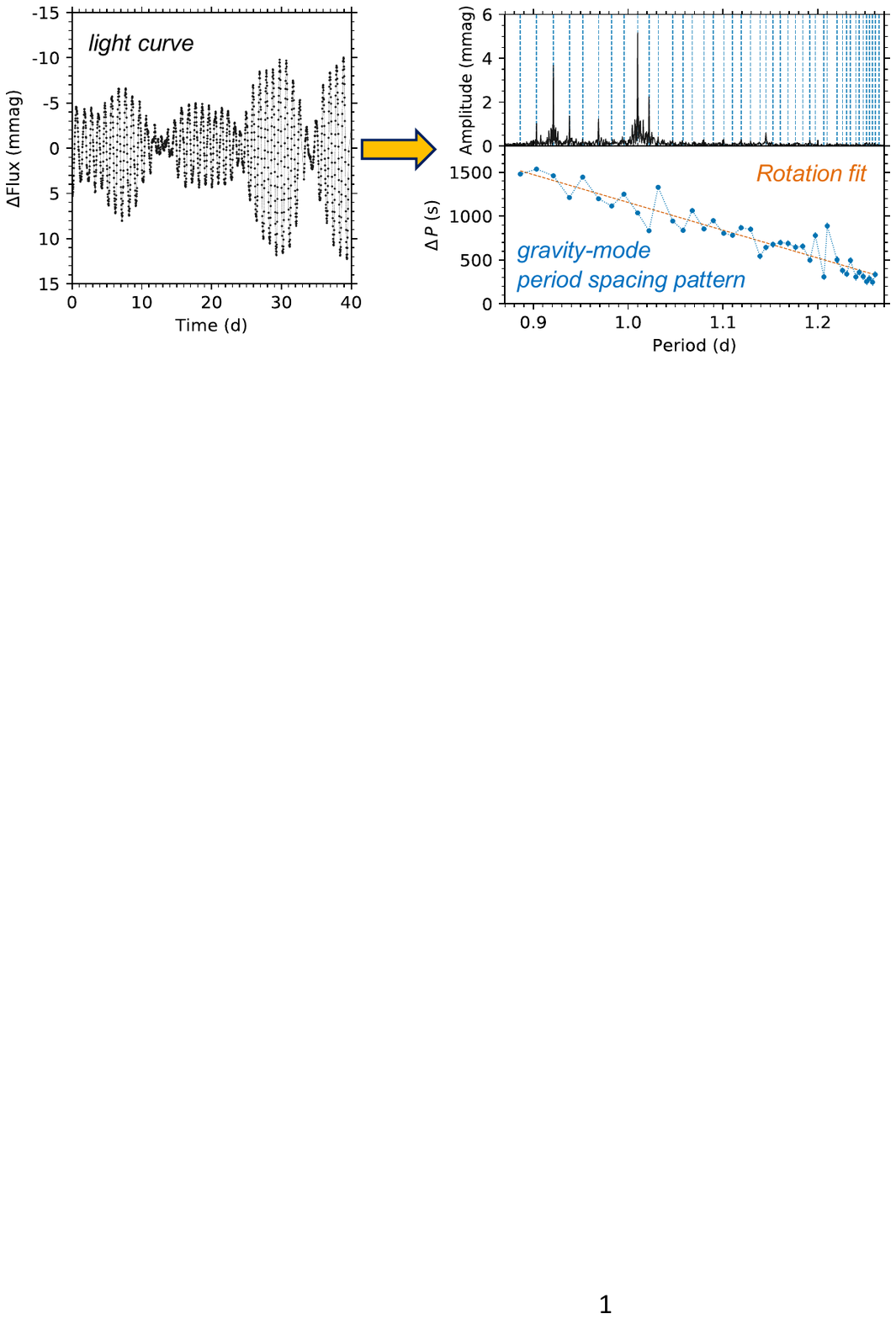}      
\caption{Schematic of the methodology of constraining interior rotation using g~modes. {\it Left:} Light curve of the Slowly pulsating B (SPB) star KIC~3459297 pulsating in low-frequency g~modes (cf. \citealt{Papics2017a}). {\it Right:} amplitude spectrum and period-spacing pattern of the prograde g-mode pulsations in the light curve are shown as the top and bottom panels, respectively. \citet{Papics2017a} determined $f_{\rm rot} = 0.63 \pm 0.04$~d$^{-1}$ from the g-mode period spacing pattern for KIC~3459297.}
\label{figure: pattern}
\end{figure}

Consequently, a powerful diagnostic in interpreting the oscillation spectrum of a rotating star pulsating in g~modes is its period spacing pattern, which is defined as the period differences, $\Delta P$, of consecutive radial order ($n$) gravity modes of the same angular degree ($\ell$) and azimuthal order ($m$) as a function of the pulsation mode period, $P$. An example of an observed period spacing pattern for a series of prograde dipole g~modes in the star KIC~3459297 \citep{Papics2017a} is shown in Fig.~\ref{figure: pattern}, in which a fit to the g-mode period spacing pattern reveals the near-core rotation rate --- see \citet{VanReeth2016a}, \citet{Ouazzani2017a} and \citet{Papics2017a} for the application of this technique. Under the asymptotic approximation, gravity modes in a non-rotating, chemically homogenous star are equally spaced in period (cf. Equation~\ref{equation: periods}), yet rotation and a chemical gradient left behind from nuclear burning within a receding convective core introduce perturbations in the form of a “tilt” and “dips”, respectively \citep{Miglio2008b, Bouabid2013}. Higher rotation rates induce a larger “tilt” with the gradient being negative for prograde modes and positive for retrograde modes in the inertial frame.

The commonly-used and mathematically appropriate approach to including rotation in the numerical computation of pulsation mode frequencies is the use of the Traditional Approximation for Rotation (TAR; \citealt{Eckart1960, Lee_U_1987b, Lee_U_1987c, Bildsten1996, Townsend2003a}). Such a treatment of the Coriolis force is necessary for g-mode pulsators if $2\Omega/\omega \gtrsim 1$ \citep{Aerts2018b}. This is because high-radial order gravity modes in moderately and rapidly-rotating stars are in the gravito-inertial regime \citep{Aerts2019b}. Within the formalism of the TAR, the horizontal component of the rotation vector is ignored, which is a reasonable assumption for gravito-inertial modes in main sequence stars given that the Lagrangian displacement vector is predominantly horizontal. The differential equations for non-radial pulsations in rotating stars are almost equivalent to those of non-rotating stars (using the Cowling approximation) if $\ell(\ell + 1)$ is replaced by the eigenvalue of the Laplace tidal equation, $\lambda$. Hence, the mathematical framework of the TAR allows the asymptotic approximation to be used for high-radial g~modes in rotating stars (see \citealt{Lee_U_1997}, \citealt{Townsend2003a} and \citealt{Townsend2003b}). Today, the TAR has been used to probe the impact of rotation on g-mode period spacing patterns both theoretically (e.g. \citealt{Bouabid2013}) and observationally (e.g. \citealt{VanReeth2016a}), and has been extended by \citet{Mathis2009d} and \citet{Mathis2019b} to take into account differential rotation and the slight deformation of stars, respectively. The TAR is implemented within the state-of-the-art pulsation code {\sc gyre} \citep{Townsend2013b, Townsend2018a} and has been used by various observational studies to probe (differential) rotation inside g-mode pulsators \citep{VanReeth2016a, VanReeth2018a}. We refer the reader to \citet{Aerts2020a**} for a detailed discussion of the TAR and its application to pulsating stars.

\vspace{0.5cm}


\subsubsection{Interior mixing}
\label{subsubsection: g-mode mixing}

Since massive stars have convective cores and radiative envelopes during the main sequence, the physics of convection and convective-boundary mixing is crucial in determining their core masses and evolution \citep{Kippenhahn_BOOK}. The mixing profile at the interface of convective and radiative regions, and the mixing profile within the envelope directly impact the amount of hydrogen available for nuclear burning. Mixing at the boundary of convective regions, such as near the convective core in a main sequence star, is typically implemented as “overshooting” in numerical codes and expressed in terms of the local pressure scale height \citep{Freytag1996b, Herwig2000c}. This is predicated on the non-zero inertia of convective bubbles at a convective boundary causing them to overshoot into a radiative layer. In massive stars, the overshooting of the convective core (also known as convective-boundary mixing) entrains hydrogen from the envelope into the core resulting in a longer main sequence lifetime and a larger helium core mass \citep{Pedersen2018a, Michielsen2019a}. This has a direct impact on the characteristic g-mode period, $\Pi_{0}$, of main-sequence stars with convective cores \citep{Mombarg2019a}.

A non-zero amount of convective core overshooting is necessary when interpreting pulsations in massive stars using 1D stellar evolution codes \citep{Dupret2004b, Briquet2007e, Daszy2013b}. Yet, the amount and shape of convective-boundary mixing remains largely unconstrained for such stars. Two examples of typical shapes of convective-boundary mixing profiles currently implemented in evolution codes include a “step overshoot” and an “exponential overshoot” (see e.g. \citealt{Herwig2000c, Paxton2015}). Typically, these two prescriptions in the shape of convective core overshooting are referred to as $\alpha_{\rm ov}$ and $f_{\rm ov}$, respectively, in the literature and differ approximately by a factor of 10-12 \citep{Moravveji2015b}. However, it is only recently that asteroseismology has demonstrated the potential to discriminate them in observations using g-mode period spacing patterns \citep{Moravveji2015b, Moravveji2016b, Pedersen2018a}. Moreover, there is considerable ongoing work using 3D hydrodynamical simulations \citep{Augustson2019a} and g-mode pulsations to probe the temperature gradient within an overshooting layer and ascertain if it is adiabatic, radiative, or intermediate between the two \citep{Michielsen2019a}.

In addition to the need for convective-boundary mixing in massive stars, the origin of mixing within their radiative envelopes is also unconstrained within evolutionary models. Direct evidence for needing increased envelope mixing comes from enhanced surface nitrogen abundances in massive stars \citep{Hunter_I_2009a, Brott2011b}. Since nitrogen is a by-product of the CNO cycle of nuclear fusion in a massive star \citep{Kippenhahn_BOOK}, an efficient mixing mechanism in the stellar envelope must bring it to the surface. Rotationally-induced mixing has been proposed as a possible mechanism \citep{Maeder2000a}, but it is currently unable to explain observed surface nitrogen abundances in slowly-rotating massive stars in the Milky Way and low-metallicity Large Magellanic Cloud (LMC) galaxies \citep{Hunter_I_2008b, Brott2011b}. Nor can rotational mixing fully explain surface abundances in massive overcontact systems \citep{Abdul-Masih2019a, Abdul-Masih2020a}. Furthermore, there was no statistically-significant relationship between the observed rotation and surface nitrogen abundance in a sample of galactic massive stars studied by \citet{Aerts2014a}. In fact, the only robust correlation with surface nitrogen abundance in the sample was the dominant pulsation frequency \citep{Aerts2014a}, which suggests that pulsations play a significant role in determining the mixing properties within the interiors of massive stars.

\vspace{0.5cm}


\subsubsection{Period spacing patterns as probes of interior rotation and mixing}
\label{subsubsection: period spacing patterns}

As previously illustrated in Figure~\ref{figure: pattern}, an observed g-mode period spacing pattern provides direct insight of the interior rotation rate of a star. Such patterns also allow the amount of interior mixing in terms of both convective core overshooting and envelope mixing to be determined. Since massive stars have a receding core whilst on the main sequence, they develop a chemical gradient in the near-core region as they evolve. Gravity modes are particularly sensitive to this molecular weight ($\mu$) gradient and in the absence of large amounts of internal mixing this leads to mode trapping \citep{Aerts2020a**}. Thus the g~modes get trapped which leads to ``dips'' in the g-mode period spacing pattern on top of the overall ``tilt'' caused by rotation (cf. Figure~\ref{figure: pattern}).

\begin{figure}[ht!]
\centering
\includegraphics[width=0.95\textwidth,clip]{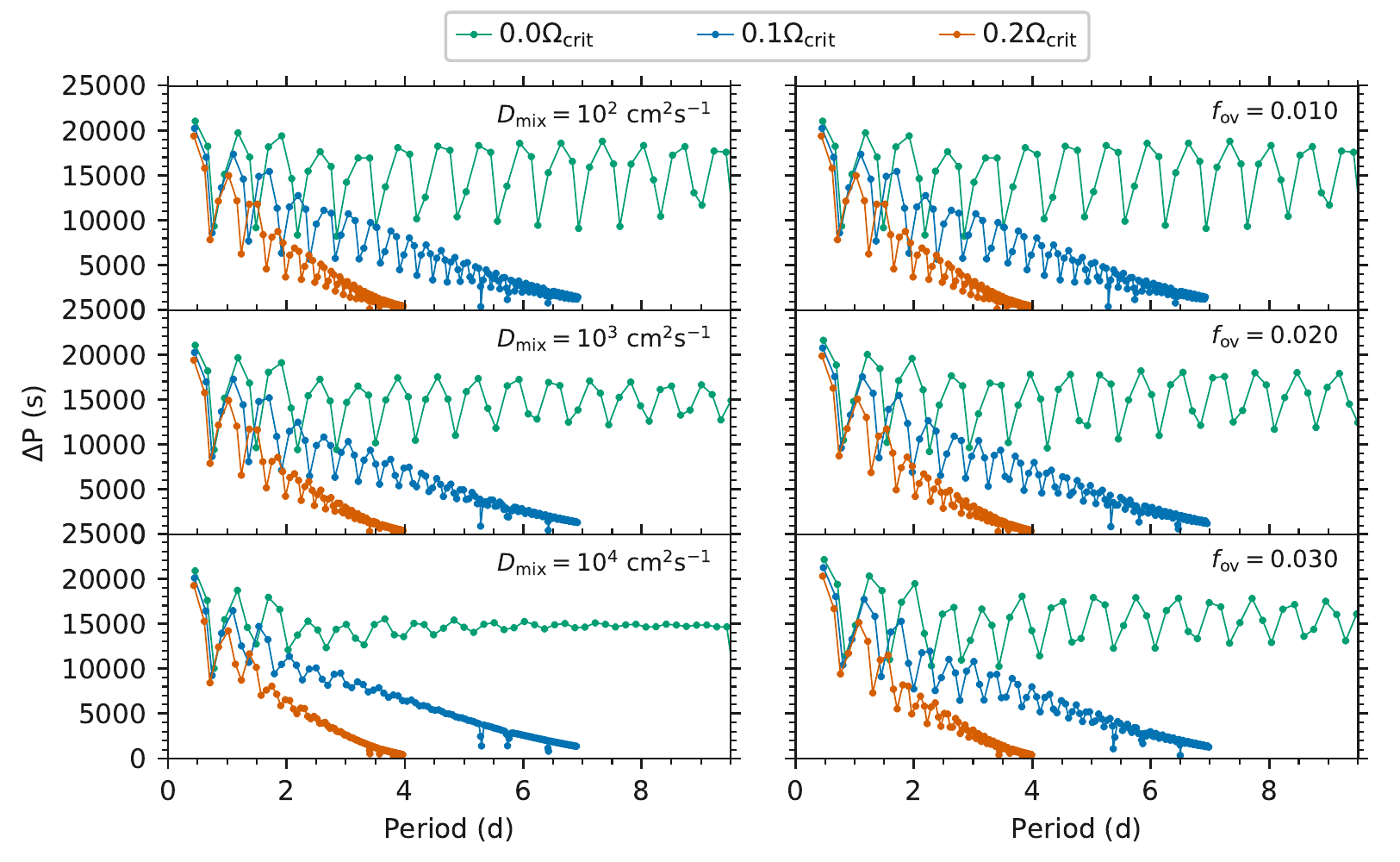}      
\caption{Theoretical gravity-mode period spacing patterns for prograde dipole modes of a 12~M$_{\odot}$ star about halfway through the main sequence (i.e. $X_{\rm c} = 0.4$). The left and right columns are for different envelope mixing ($D_{\rm mix}$) in cm$^{2}$\,s$^{-1}$, and exponential convective-boundary mixing ($f_{\rm ov}$) expressed in local pressure scale heights, respectively, calculated using the {\sc mesa} stellar evolution code \citep{Paxton2019}. For each panel, three different rotation rates expressed as a fraction of the critical rotation rate, $\Omega_{\rm crit}$, calculated using the Traditional Approximation for Rotation (TAR) using the {\sc gyre} pulsation code \citep{Townsend2013b} are shown.}
\label{figure: GYRE}
\end{figure}

An illustration of the effect of different amounts of interior mixing and rotation on the g-mode period spacing patterns of prograde dipole modes in a $12$~M$_{\odot}$ star about halfway through the main sequence is shown in Figure~\ref{figure: GYRE}. In the left column of Figure~\ref{figure: GYRE}, the effect of increasing the amount of envelope mixing (denoted by $D_{\rm mix}$) is shown going from top to bottom. Whereas in the right column, the effect of increasing the amount of convective core overshooting (denoted by $f_{\rm ov}$) is shown from top to bottom. Such values of $D_{\rm mix}$ and $f_{\rm ov}$ represent the range of values typically found by asteroseismic studies of stars with convective cores. Clearly, even for moderate values of envelope mixing (i.e. $D_{\rm mix} \simeq 10^{4}$~cm$^{2}$\,s$^{-1}$), the presence of dips in the g-mode period spacing pattern are strongly diminished, as shown in the bottom-left panel of Figure~\ref{figure: GYRE}. Thus, the observed presence of strong dips in g-mode period spacing patterns already places an upper limit on the amount of envelope mixing possible in such stars. In all panels of Figure~\ref{figure: GYRE}, three different rotation rates are shown in green, blue and red, demonstrating the significant affect of rotation for g~modes, which correspond approximately to rotation frequencies of 0.0, 0.1 and 0.2~d$^{-1}$, respectively. The effect of rotation was calculated assuming rigid rotation using the TAR implemented in the {\sc gyre} pulsation code \citep{Townsend2013b,Townsend2018a}. 

Yet, it is possible that only some, or even none, of the pulsation modes shown in Fig.~\ref{figure: GYRE} are observed in massive stars, since the excitation of a given pulsation mode depends on stellar parameters, including mass, age, metallicity. The example using a 12~M$_{\odot}$ shown in Fig.~\ref{figure: GYRE} does not include any predictions of mode excitation. Nevertheless, Fig.~\ref{figure: GYRE} serves as a schematic example of similar behaviour for any main sequence star born with a convective core. In summary, the morphology of an observed gravity-mode period spacing pattern facilitates mode identification and offers a direct measurement of the near-core rotation and chemical mixing within a star. In practice, asteroseismology of g~modes requires long-term and high-precision time series (space) photometry to extract g-mode period spacing patterns. These patterns consequently allows one to measure the interior rotation and constrain the envelope mixing of the host star, and the corresponding characteristic g-mode period, $\Pi_{0}$, which places constraints on its mass and age. From a large and multi-dimensional grid of stellar structure models covering the possible values of mass, age, and interior mixing parameterised by $f_{\rm ov}$ and $D_{\rm mix}$, a quantitative comparison of observed and theoretical gravity-mode period spacing patterns facilitates asteroseismology to derive the interior properties of massive stars \citep{Aerts2018b}.

\vspace{0.5cm}


\subsection{Instability domains of high-mass stars}
\label{subsection: instability regions}

The common interior structures of massive stars includes a convective core and radiative envelope, with pulsations in these stars being driven by the $\kappa$-mechanism operating within the local opacity enhancements caused by the Z-bump associated with iron-peak elements in their near-surface layers \citep{Dziembowski1993e, Dziembowski1993f, Gautschy1993a, Pamyat1999b, Miglio2007a}. The depth of the Z-bump at $T \simeq 200\,000$~K depends on the effective temperature of the star and in turn defines an upper and lower temperature boundary for the instability region of the $\kappa$-mechanism in the HR~diagram, which are sometimes referred to the blue and red edges of an instability region, respectively. In addition to the mass, radius and effective temperature of a star, which define its thermal structure, the metallicity and choice of opacity table in models are also important parameters \citep{Dziembowski2008, Paxton2015, Walczak2015, Daszy2017a}. Since the $\kappa$-mechanism operates in the Z-bump, it requires a sufficiently-large opacity enhancement to block radiation and excite coherent pulsation modes. This is supported by theoretical models of pulsation excitation, and observations which indicate a dearth of massive pulsators in low-metallicity environments such as the LMC galaxy with $Z \simeq 0.5\,Z_{\odot}$ (see e.g. \citealt{Salmon_S_2012}). 

Furthermore, rotation plays an important role in defining the instability regions of massive stars. From an observational perspective, moderate and fast rotation distorts the spherical symmetry of a star. This has significant implications for the spectroscopic determination of atmospheric parameters such as the effective temperature and surface gravity, as these parameters are significantly affected by gravity darkening \citep{vonZeipel1924b, Townsend2004b, EspinosaLara2011}. From a more theoretical perspective, the distorted spherical symmetry of rapidly-rotating stars impacts the applicability of using 1D models, and because phenomena associated with rapid rotation such as rotationally-induced mixing, can significantly influence evolutionary tracks in the HR~diagram \citep{Maeder2000a, Maeder_rotation_BOOK, Lovekin2020a}. Moreover, as discussed in section~\ref{subsubsection: g-mode rotation}, the Coriolis force perturbs the pulsation frequencies of a rotating star and consequently also the expected parameter range of instability regions in the HR~diagram \citep{Townsend2005b, Bouabid2013, Szewczuk2017a}.

The calculation of instability regions for stars requires non-adiabatic calculations, and specifically the calculation of a pulsation mode's growth rate \citep{UNNO_BOOK, ASTERO_BOOK}. Following the laws of thermodynamics, heat-driven pulsation modes require that heat is gained in phase with compression during a pulsation cycle. A non-adiabatic pulsation calculation yields a mode's eigenfrequency, and its imaginary component yields the growth rate. For heat-driven modes excited by the $\kappa$-mechanism, the growth rate is a positive quantity for modes that are effectively excited and negative for modes that are damped \citep{UNNO_BOOK, ASTERO_BOOK}. The instability regions of pulsations in massive stars for different masses, ages, metallicities and rotation rates can be readily calculated for early-type stars by means of stellar structure and evolution codes, such as {\sc mesa} \citep{Paxton2011, Paxton2013, Paxton2015, Paxton2018, Paxton2019}, when coupled to non-adiabatic stellar pulsation codes, such as {\sc gyre} \citep{Townsend2013b,Townsend2018a}. We refer the reader to \citet{Moravveji2016a}, \citet{Godart2017} and \citet{Szewczuk2017a} for instability regions of main-sequence massive stars calculated with and without rotation, and to \citet{Daszy2013c} and \citet{Ostrowski2015b} for instability calculations for post-main sequence massive stars.

Amongst the early-type stars, there are two main groups of stars that pulsate in coherent pulsation modes excited by the $\kappa$-mechanism: the $\beta$~Cephei ($\beta$~Cep) stars and the Slowly Pulsating B (SPB) stars, which together span the approximate mass range from 3 to 25~M$_{\odot}$. Although not traditionally classified as a distinct pulsator group amongst massive stars, the pulsating Be stars are also discussed in this section for completeness. Pulsations in more evolved and/or more massive stars have also been detected, such as those in periodically variable supergiant (PVSG) stars \citep{ASTERO_BOOK}. However, they are not included here since there are currently very few asteroseismic studies of these objects. We refer the reader to \citet{Saio2006b} and \citet{Ostrowski2017a} for insightful work on the variability of such stars.

\vspace{0.5cm}


\subsubsection{$\beta$~Cephei stars}

The $\beta$~Cephei ($\beta$~Cep) stars are Population~I stars with spectral types ranging from late O to early B on the main sequence, and have birth masses larger than some 8~M$_{\odot}$ and up to 25~M$_{\odot}$. They pulsate in low-radial order g and p~modes excited by the $\kappa$-mechanism operating in the Z-bump \citep{Dziembowski1993e}, and have pulsation periods that range between about 2 and 8~hr \citep{Stankov2005, Pigulski2008a, Pigulski2008b, ASTERO_BOOK}. Most $\beta$~Cep stars are dwarf stars, making them likely main sequence stars, although a significant fraction are giants or supergiants. Our understanding of the driving of low-radial order g~modes in high-mass $\beta$~Cep stars remains somewhat elusive \citep{Handler2004b, Handler2017a, Aerts2004b}, since a substantial overabundance of iron and nickel in the Z-bump is typically needed for the $\kappa$-mechanism to be efficient at exciting g~modes in such stars \citep{Pamyat2004, Moravveji2016a, Daszy2017a}. Moreover, in addition to the mode excitation by the $\kappa$-mechanism, $\beta$~Cep stars have been shown to exhibit non-linear mode excitation \citep{Degroote2009a} and stochastically-excited pulsation modes \citep{Belkacem2009b, Belkacem2010a, Degroote2010b}, which demonstrate that multiple pulsation driving mechanisms exist in these stars.

\vspace{0.5cm}


\subsubsection{Slowly Pulsating B stars}

The Slowly Pulsating B (SPB) stars are the lower-mass counterparts of the $\beta$~Cep stars, with the original definition of this type of variable star made by \citet{Waelkens1991c} although individual examples of SPB stars were known previously (e.g. 53~Persei; \citealt{Smith_M_1978c}). The SPB stars are Population~I stars with spectral types that range from B3 to B9 on the main sequence, thus they have birth masses between approximately 3 and 8~M$_{\odot}$ \citep{ASTERO_BOOK}. They pulsate in high-radial order and predominantly prograde dipole g~modes excited by the $\kappa$-mechanism operating in the Z-bump \citep{Dziembowski1993f, Gautschy1993a} and have pulsation periods that range between a few days and several hours \citep{Waelkens1998c, Aerts1999a, DeCat2002c}. 

\vspace{0.5cm}


\subsubsection{Pulsating Be stars}

A subset of approximately 20\% of non-supergiant massive stars are classified as Be stars \citep{Porter_J_2003b}. This group comprises stars that have shown Balmer lines in emission on at least one occasion, since such emission lines are known to be transient \citep{Zorec1997b, Neiner2011c, Rivinius2013c}. The Be stars are rapid rotators with circumstellar decretion disks, and their near-critical rotation rates are thought to be related to their evolutionary history. More specifically, Be stars may have accreted mass from a companion or are the result of a stellar merger, which is supported by the relative rarity of Be stars with main-sequence companions \citep{Bodensteiner2020b*}. Many Be stars show evidence of pulsations and experience outbursts of material thought to be driven by pulsations \citep{Rivinius2003, Huat2009c, Kurtz2015b}. The pulsational behaviour of Be stars is quite diverse, with such stars showing coherent g~modes and stochastically excited gravito-interial waves, with the amplitude of their pulsations being connected to whether the star is mid-outburst or in quiescence \citep{Kambe1993a, Porter_J_2003b, Neiner2009d, Neiner2012d, Baade2016a}. Given the diverse variability seen in Be stars, it remains unclear if a single excitation mechanism is unanimously responsible for exciting pulsations in Be stars, with their fast rotation being their main common characteristic \citep{Porter_J_2003b, Townsend2004b}.

\vspace{0.5cm}


\section{The space photometry revolution}
\label{section: space photometry}

The successful application of asteroseismology requires long-term, continuous and high-precision time series data to resolve individual pulsation mode frequencies, and perform unambiguous mode identification. As discussed in sections~\ref{subsection: pressure modes} and \ref{subsubsection: period spacing patterns}, mode identification using the continuous photometry provided by space telescopes can be readily achieved using rotationally-split modes and period spacing patterns, respectively. However, prior to space telescope missions different techniques were more common. Nevertheless, once mode identification has been achieved, a quantitative comparison of observed pulsation mode frequencies and those predicted by theoretical models in addition to atmospheric constraints such as the effective temperature and metallicity from spectroscopy reveals the physics that best represents the observed star --- a fitting process known as forward seismic modelling \citep{Aerts2018b}.

\vspace{0.5cm}


\subsection{Prior to space telescopes}
\label{subsection: before}

Prior to the near-continuous photometry provided by space telescopes, pulsation modes in massive stars were identified by means high-resolution and high-cadence spectroscopic and/or photometric time series data assembled using ground-based telescopes (e.g. \citealt{Smith_M_1977, Waelkens1991c, DeCat2005}). In the early days of massive star asteroseismology, mode identification via spectroscopic line profile variations (LPVs) targeted the silicon triplet at 4560~\AA{} in slowly-rotating $\beta$~Cep stars \citep{Gies1988a, Aerts1992b, Aerts1993, Aerts1994b, Aerts1994c, Telting1997a, Telting1997c, Uytterhoeven2004b, Uytterhoeven2005a}, whereas for SPB stars the silicon doublet at 4130~\AA{} was also useful \citep{Aerts1999c, Aerts2003a}. A spectral resolving power of at least 50\,000 and a very high signal-to-noise being preferable. In fast rotating B stars, such as Be stars, these silicon multiplets can be blended and one must target isolated lines such as the helium~I 6678~\AA{} line (e.g. \citealt{Balona1999b, Balona1999f, Stefl2003b, Maintz2003}). Balmer lines are not suitable for LPV studies as they are less sensitive to the radial and non-radial velocities of pulsations since they are dominated by intrinsic (i.e. Stark) broadening. For a complete discussion of mode identification using spectroscopic time series, we refer the reader to \citet{ASTERO_BOOK}.

An alternative and complementary methodology to using LPVs in the identification of pulsation modes in massive stars is to use amplitude ratios from multi-colour photometry. Originally devised by \citet{Watson1988} and \citet{Heynderickx1994b}, this approach was particularly effective when applied to $\beta$~Cep stars owing to their relatively high amplitude and short-period pulsations. We refer the reader to \citet{Shobbrook2006a}, \citet{Handler2012a} and \citet{Handler2017a} for examples of this technique. The analysis of ground-based photometry and/or spectroscopy has been successful in demonstrating the importance of constraining parameters such as convective core overshooting, rotation and metallicity in massive star evolution \citep{Aerts2003d, Handler2004b, Handler2006a, Briquet2007e, Daszy2013b, Szewczuk2015c}.

\vspace{0.5cm}


\subsection{Early space missions}
\label{subsection: space}

When the era of space telescopes dawned, the revolution of asteroseismology for stars across the HR~diagram truly began. Space telescopes not only provide long-term and near-continuous observations of multiple stars simultaneously, but the typical precision of space-based photometry is at least two orders of magnitude better than what is possible from the ground. Such early space missions, which were not designed for asteroseismic surveys but nonetheless were extremely useful, included the {\it Hipparcos} mission \citep{vanLeeuwen1997c}, which discovered hundreds of pulsating B~stars \citep{Waelkens1998a, Aerts2006f, Lefevre2009c}.  The drastically improved time series data allowed low-amplitude pulsation modes to be detected in stars previously believed to be constant, and yielded precise pulsation frequencies in multi-periodic pulsators that are dominated by complex beating patterns \citep{Koen2002c}. 

The first space telescope dedicated to asteroseismology was the MOST mission, which was launched in 2003 \citep{Walker2003}. The MOST spacecraft may have been a small telescope, but it detected pulsations in a wide variety of different types of variable stars. Among the massive stars, MOST studied SPB, $\beta$~Cep and Be stars \citep{Walker2005b, Aerts2006d, Aerts2006e, Saio2007a, Saio2007c, Cameron2008}, including rare instances of $\beta$~Cep stars in eclipsing binary systems \citep{Desmet2009a, Desmet2009c}. Simultaneous MOST photometry and ground-based spectroscopy proved vital in understanding and modelling the first massive star exhibiting hybrid p- and g-mode behaviour, $\gamma$~Peg \citep{Handler2009b, Walczak2013b}. Furthermore, MOST detected pulsations in blue supergiants, such as Rigel \citep{Saio2006b, Moravveji2012a}, and interestingly it also detected pulsations in some Wolf-Rayet stars \citep{Lefevre2005c, Moffat2008d} but the absence of pulsations in others \citep{Moffat2008c}. Although the time series photometry assembled by the MOST mission was limited in length, it served as a valuable proof-of-concept exercise demonstrating the power of massive star asteroseismology using space telescopes compared to facilities on the ground.

The next major milestone in the space photometry revolution was the French-led CoRoT mission launched in 2006 \citep{Auvergne2009}. The CoRoT spacecraft was a combined planet hunting and asteroseismology focussed mission, which delivered short-cadence (i.e. 32~sec) time series photometry for different fields of view across the sky for up to 150~d \citep{Baglin2009a}. Over the mission duration, CoRoT contributed a wealth of information concerning variability in massive stars, with some aspects remaining unchallenged in quality and impact more than a decade later. The CoRoT fields of view were optimised to contain hundreds of candidate pulsating B~stars \citep{Degroote2009a}, which included $\beta$~Cep, SPB and Be stars \citep{Degroote2009a, Degroote2013, Briquet2011, Papics2011, Neiner2012d, Aerts2011, Aerts2019a}. The mission was a great success for massive star variability studies, with CoRoT detecting regular frequency spacings in the O8.5\,V star HD~46149 \citep{Degroote2010b}, and providing firm confirmation that massive stars can pulsate in both p- and g-mode frequencies \citep{Degroote2012b}. Furthermore, CoRoT led to the first discovery of deviations from a constant period spacing in the B3\,V star HD~50230 \citep{Degroote2010a}. In addition to high-precision asteroseismology of targeted $\beta$~Cep stars \citep{Aerts2011, Aerts2019b}, CoRoT also discovered stochastic variability caused by pulsations in massive star photospheres. This includes stochastic non-radial pulsations in B~stars \citep{Belkacem2009b, Neiner2012d} and stochastic low-frequency variability in the three O stars HD~46223, HD~46150, and HD~46966 \citep{Blomme2011b, Bowman2019a}.

Of all the space photometry missions available for asteroseismology, the BRITE-constellation of nanosatellites is ranked amongst the highest for providing excellent asteroseismic returns considering its budget and etendue \citep{Weiss2014, Pablo2016}. The constellation of nanosatellites from the collaboration of Austria, Poland and Canada, was originally launched in 2013 and have provided long-term time series photometry of some of the brightest stars in the sky. This includes $\beta$~Cep, SPB and Be stars \citep{Pigulski2016, Baade2016a, Handler2017a, Daszy2017a, Walczak2019a}, but also pulsating stars in multiple systems \citep{Kallinger2017a, Pablo2017a, Pablo2019} and stochastic variability in O~supergiants and Wolf-Rayet stars \citep{Buysschaert2017a, Rami2018a, Rami2018b, Rami2019}. The multi-colour and long-term photometry of BRITE and its observing strategy are naturally complementary to combing BRITE data with simultaneous ground-based photometry and/or spectroscopy (e.g. \citealt{Handler2017a}).

\vspace{0.5cm}
\vspace{0.5cm}


\subsection{The {\it Kepler} and K2 missions}
\label{subsection: Kepler}

Perhaps the most famous of all space telescopes providing time series photometry, the {\it Kepler} space telescope was launched in 2009 and had a primary goal of finding Earth-like planets orbiting Sun-like stars \citep{Borucki2010, Koch2010}. Although by the end of the nominal {\it Kepler} mission, the 4-yr light curves of more than 200\,000 stars proved invaluable for asteroseismology as well as exoplanet studies. Massive stars ($> 8$~M$_{\odot}$) were purposefully avoided by the {\it Kepler} mission since they are bright and typically saturated the CCDs, although dozens of SPB stars and rotationally-variable B~stars were discovered \citep{McNamara2012, Balona2015d}. Since the end of the nominal 4-yr {\it Kepler} mission, the extremely high precision light curves have been used to discover and analyse rotationally-split modes and period spacing patterns of prograde dipole g~modes in dozens of SPB stars \citep{Papics2014, Papics2015, Papics2017a}. Such asteroseismic studies have revealed the interior rotation rates, convective core overshooting and mixing in main-sequence B stars \citep{Moravveji2015b, Moravveji2016b, Szewczuk2018a}. Despite massive stars not being included in the {\it Kepler} field of view, ingenious techniques to extract the light curves of nearby massive stars have been successful \citep{Pope2016a, Pope2019b}, which includes the scattered-light variability of the O9.5\,Iab star HD~188209 \citep{Aerts2017a}.

After approximately 4~yr the {\it Kepler} spacecraft lost a second vital reaction wheel, which meant that the field of view could not be maintained without a significant expenditure of fuel. A new mission, K2: {\it Kepler}'s second light, was devised by NASA, which consisted of 80-d campaigns pointing in the direction of the ecliptic \citep{Howell2014}. Since the K2 campaign fields included young star-forming regions, a plethora of massive stars were available to study using K2 light curves. The first proof-of-concept of O-star asteroseismology was demonstrated by \citet{Buysschaert2015}, which acted as an important proof-of-concept for massive star variability. The K2 mission allowed the rotation, pulsation and binary properties of the seven sisters of the Pleiades to be studied in detail for the first time thanks to its long time base and high precision \citep{White2017b}. Since the end of the K2 mission in 2018, these high-precision data have revealed dozens of previously unknown $\beta$~Cep and SPB stars \citep{Pope2016a, Pope2019c, Burssens2019a} and ubiquitous stochastic low-frequency variability in the photospheres of massive stars \citep{Bowman2019b}.

\vspace{0.5cm}


\subsection{The TESS mission}
\label{subsection: TESS}

The ongoing Transiting Exoplanet Survey Satellite (TESS; \citealt{Ricker2015}) is currently providing high-precision and short cadence (i.e. 2~min) observations for hundreds of thousands of stars across the sky. Each ecliptic hemisphere ($|b| > 6$~degrees) is divided into 13 sectors which are each observed for up to 28~d. However, there is overlap of the observational sectors near the ecliptic poles, such that stars within the TESS continuous viewing zones (CVZ) have uninterrupted light curves spanning 1~yr. Such an observing strategy is optimised to find transiting exoplanets orbiting bright stars across the sky, but TESS data are also extremely valuable for massive star asteroseismology. Amongst the earliest studies of massive star variability as viewed by TESS are those by \citet{Handler2019a}, \citet{Bowman2019b} and \citet{Pedersen2019a} in which the diverse variability of massive stars is demonstrated using a total sample of more than 200 massive stars. The study by \citet{Handler2019a} considers one of the first $\beta$~Cep stars observed by the TESS mission, which revealed it to be multi-periodic and subsequent asteroseismic modelling indicated it was a runaway star because of its inferred age.

\begin{figure}[ht!]
\centering
\includegraphics[width=0.5\textwidth,clip]{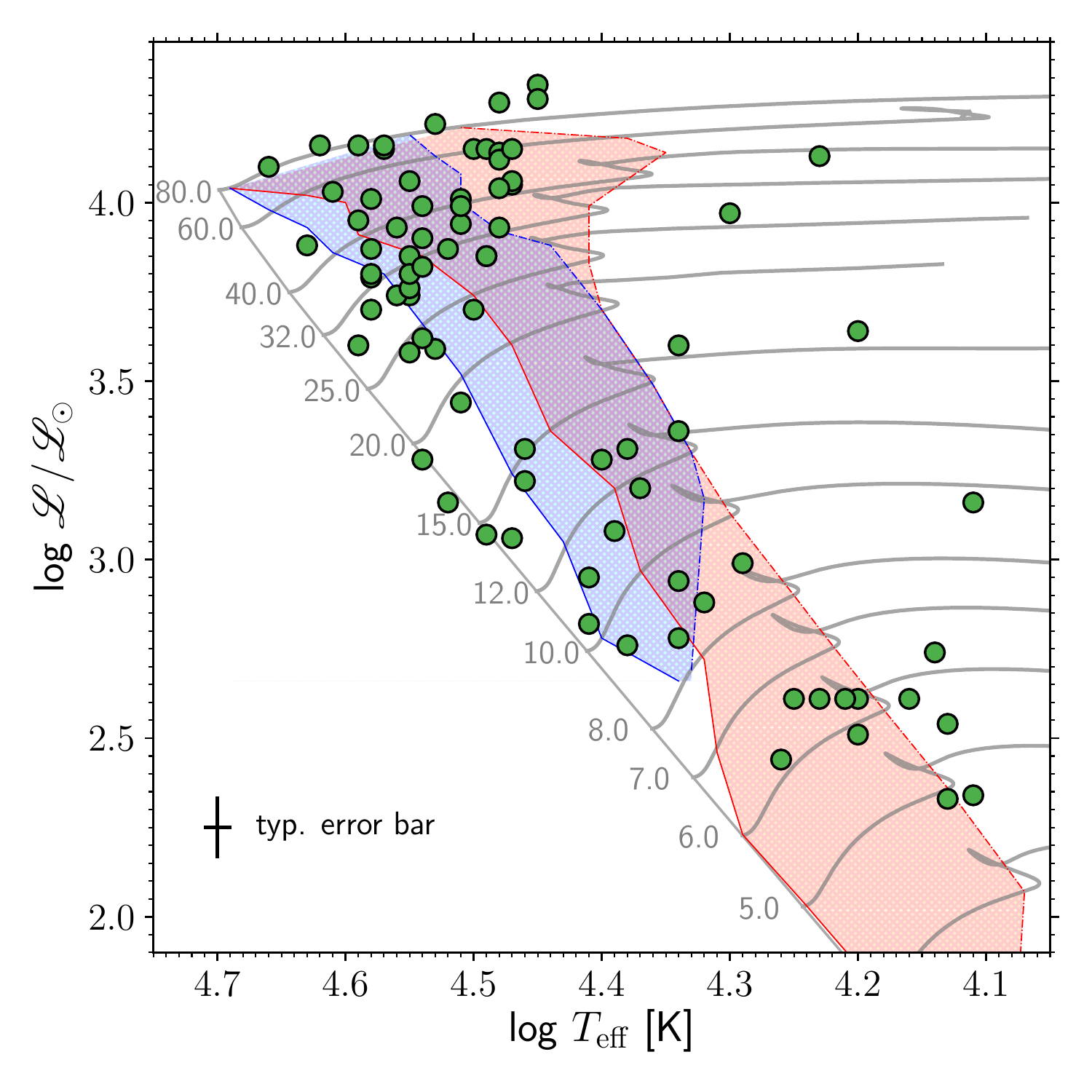}      
\caption{Spectroscopic HR~diagram of the pulsating massive stars observed by TESS in its sectors 1--13, which have spectroscopic parameters derived by high-resolution spectroscopy, are shown as green circles. Note that the ordinate axis shows spectroscopic luminosity such that $\mathscr{L} := T_{\rm eff}^{4} / g$ \citep{Langer2014a}. Red and blue hatched regions denote the theoretical instability regions of g and p~modes, respectively, for main sequence stars. Non-rotating evolutionary tracks at solar metallicity (in units of M$_{\odot}$) are shown as solid grey lines, and a typical spectroscopic error bar is shown in the bottom-left corner. Figure adapted from \citet{Burssens2020a}, their figure~3.}
\label{figure: TESS HRD}
\end{figure}

Since the first 13 sectors of TESS data in the southern ecliptic hemisphere have become available, \citet{Burssens2020a} have completed a census of massive star variability using TESS data, and coupled this to high-resolution spectroscopy from the IACOB \citep{Simon-Diaz2011d, Simon-Diaz2014a, Simon-Diaz2015c} and OWN \citep{Barba2010, Barba2014, Barba2017} surveys. This allowed them to place more than 100 pulsating massive stars in the spectroscopic HR~diagram. A summary of the variability catalogue by \citet{Burssens2020a} is shown in the spectroscopic HR~diagram in Figure~\ref{figure: TESS HRD}, in which the observed stars are shown as green circles overlaid on top of evolutionary tracks (grey lines) and theoretical instability regions for p and g~modes (blue and red hatched regions, respectively), which were calculated using the {\sc mesa} evolution code \citep{Paxton2019} and the {\sc gyre} pulsation code \citep{Townsend2013b}. Although there is general agreement between the observed location of pulsating stars and the predicted instability regions, the study by \citet{Burssens2020a} clearly demonstrates that our understanding of the variability mechanisms in massive stars is far from complete given the overall distribution of stars within the HR~diagram.


\section{Empirical constraints on massive star interiors}
\label{section: constraints}

In the previous section, it was described how early asteroseismic studies of massive star interiors were predominantly based on ground-based campaigns. The acquisition of these necessary data was painstakingly complex and required substantial dedication from all those involved, since it is non-trivial to collect and analyse such fragmented time series. However, despite the complications some aspects of these ground-based studies remain unrivalled to this day owing to the carefully-selected sample of stars and the relative exclusion of massive stars in later space missions. In particular, there remain only a handful of truly massive stars that have been undergone forward seismic modelling. In this section, some important case studies of pulsating massive stars that have studied using asteroseismology are presented. In particular the range of parameters that have been obtained via forward seismic modelling, where available, are summarised in Table~\ref{table: modelling}, but we refer the reader to the individual studies for full details.

\vspace{0.5cm}
\vspace{0.5cm}


\subsection{Ground-based studies}
\label{subsection: ground}

The first detailed study of the interior of a massive star using asteroseismology is arguably that of the slowly-rotating B3\,V star V836~Cen (HD~129929) \citep{Aerts2003d, Aerts2004c, Dupret2004b}. The analysis of the extensive ground-based data set consisting of multicolour Geneva photometry and high-resolution spectroscopy revealed HD~129929 to be a multiperiodic $\beta$~Cep star pulsating in low-radial order g and p modes \citep{Aerts2003d, Aerts2004c}. The rotationally-split multiplets in HD~129929 yielded a non-rigid interior rotation rate where the near-core region was determined to be rotating a factor of $\sim$4 times faster than the envelope. This was the first measurement of the interior rotation rate of a main-sequence star. Subsequent forward seismic modelling allowed the best-fitting mass, age, metallicity, core hydrogen mass fraction and convective core overshooting to be determined \citep{Dupret2004b}, which are provided in Table~\ref{table: modelling}.

An important example of extensive ground-based multicolour photometry and spectroscopy being used to probe the interior of a $\beta$~Cep star is that of B2\,III star $\nu$~Eri (HD~29248) \citep{Handler2002l, Handler2004b, Aerts2004b, Pamyat2004, DeRidder2004d, Ausseloos2004, Jerzykiewicz2005, Suarez2009a, Daszy2010a}. At the time, and arguably still true today, $\nu$~Eri represents one of the best studied $\beta$~Cep stars owing to the substantial data set having been assembled for the star. The spectroscopic and frequency analyses of this slowly-rotating ($v\,\sin\,i \lesssim 20$~km\,s$^{-1}$) star revealed rotationally-split low-radial order dipole g and p~modes, and possibly high-radial order g~modes \citep{Handler2004b, Aerts2004b}. Forward seismic modelling of $\nu$~Eri was unable to find a satisfactory theoretical model unless an iron enhancement throughout the star, and in particular in the Z-bump, was included in the models to explain the mode excitation \citep{Pamyat2004, Ausseloos2004}. The best-fitting parameters of $\nu$~Eri yielded a non-rigid interior rotation rate of approximately 3-5 times faster in the near-core region compared to the envelope, although this is quite uncertain given the asymmetry of the small number of rotationally-split modes available \citep{Pamyat2004, Ausseloos2004}. Later, \citet{Dziembowski2008}, \citet{Suarez2009a}, and \citet{Daszy2010a} revisited the modelling of $\nu$~Eri and investigated its interior rotation, overshooting and mode excitation properties based on different opacity tables, specifically focussing on the relative abundance of iron in the Z-bump. In particular, these asteroseismic modelling studies were able to produce a better fit to the observed pulsation mode frequencies using larger overshooting values and an enhancement of iron in the Z-bump, albeit at the expense of not necessarily re-producing the inferred location of the star in the HR~diagram. The best-fitting parameters from these studies are provided in Table~\ref{table: modelling}, although the variance in the parameters can be understood as arising from the different metallicities and opacity tables being used.

A similarly famous $\beta$~Cep star is 12~Lac (B1.5\,III; HD~214993), which is comparable in mass to $\nu$~Eri although it is somewhat more rapidly rotating with a projected rotational velocity of $30 \lesssim v\,\sin\,i \lesssim 40$~km\,s$^{-1}$ \citep{Gies1992b, Abt2002}. However, producing a satisfactory asteroseismic model for 12~Lac has been difficult owing to inconsistencies between observed and theoretically-predicted mode identifications, the consequential impact on the inferred interior rotation profile, and stellar opacity data being unable to explain the excitation of all observed pulsation mode frequencies \citep{Aerts1996b, Dziembowski1999a, Handler2006a, Dziembowski2008, Desmet2009b}. Despite these difficulties, \citet{Dziembowski2008} conclude based on the then-available observations of 12~Lac that it also exhibits an interior rotation rate of approximately 4-5 times faster in the near-core region compared to the envelope. \citet{Desmet2009b} were able to constrain the parameters of 12~Lac, but the most advanced study of 12~Lac to date was performed by \citet{Daszy2013b}, who performed mode identification, asteroseismic modelling and studied how rotation, metallicity and opacity data significantly affect the modelling results.

The slowly-rotating ($v\,\sin\,i \simeq 30$~km\,s$^{-1}$) $\beta$~Cep star $\theta$~Oph (B2\,IV; HD~157056) was also subject to intense photometric campaigns to detect, extract and identify its variability \citep{Handler2005b, Briquet2005, Daszy2009b}. The frequency spectrum of $\theta$~Oph resembled that of V836~Cen in that it also contained a single radial mode and a handful of low-radial order rotationally-split multiplets, although one difference is that $\theta$~Oph is known to be a multiple system \citep{Briquet2005}. However, contrary to previous examples of $\beta$~Cep stars discussed in this section, forward seismic modelling of $\theta$~Oph by \citet{Briquet2007e} revealed an approximately rigid interior rotation rate and a significantly larger amount of convective core overshooting with the best-fitting model yielding $\alpha_{\rm ov} = 0.44 \pm 0.07$. Such a high value of convective core overshooting is rare in the context of modern asteroseismology of stars with convective cores \citep{Aerts2019b}, but does highlight the trend that stellar models typically underestimate the mass of the convective cores in high-mass main-sequence stars.

A final example case study of forward seismic modelling of a slowly-rotating $\beta$~Cep star based on ground-based data is the B2\,IV/V star V2052~Oph (HD~163472; \citealt{Briquet2012}). The importance of V2052~Oph in this context is that it is a known magnetic star \citep{Neiner2003c, Neiner2012a}. A large-scale magnetic field is thought to suppress the near-core mixing caused by convective core overshooting in massive stars. The data set of V2052~Oph assembled and analysed by \citet{Briquet2012} included more than 1300 spectra from 10 different telescopes around the world, and the resultant frequency spectrum of included a radial mode and two prograde non-radial modes identified by means of spectroscopy. Forward seismic modelling of these identified pulsation modes yielded a relatively fast rotation rate with $v_{\rm eq} \simeq 75$~km\,s$^{-1}$, and only a small amount of convective core overshooting with $\alpha_{\rm ov} \in [0.00,0.15]$, which was concluded to be low because of the star's magnetic field \citep{Briquet2012}.

\vspace{0.5cm}


\subsection{Space-based studies}
\label{subsection: mixing}

Early space missions, such as MOST \citep{Walker2003} have targeted massive stars for the purposes of asteroseismology. However, in this review we focus on selected case studies from the BRITE, CoRoT and {\it Kepler} missions. This is because of their high photometric precision and their long observational base lines \citep{Auvergne2009, Baglin2009a, Borucki2010, Koch2010, Weiss2014}, which are necessary for successful forward seismic modelling.

The CoRoT mission provided near-continuous light curves up to 150~d in length at a cadence of 32~s, which at the time was a major revolution for asteroseismology. Early important results from the mission included the seismic modelling of the $\beta$~Cep star HD~180642 \citep{Aerts2011}, and detection of the g-mode period spacing pattern in the B3\,III SPB star HD~50230 \citep{Degroote2010a}. Such a g-mode period spacing pattern was a huge step forward in asteroseismology, as such patterns enable mode identification and allow interior rotation and mixing to be measured directly. Subsequent modelling of HD~50230 revealed it to be a mid-main sequence star with a mass of approximately 7~M$_{\odot}$, and convective core overshooting of approximately $0.2 \leq \alpha_{\rm ov} \leq 0.3$ \citep{Degroote2010a}. Later, \citet{Degroote2012b} discovered that HD~50230 is a wide-binary system, and confirmed its ultra-slow rotation rate of $v_{\rm eq} \simeq 10$~km\,s$^{-1}$ using spectroscopy, although they concluded that this does not impact the aforementioned g-mode period spacing pattern. More recently, \citet{Wu_T_2019a} have revisited the analysis of HD~50230 and conclude it to be a metal-rich hybrid pulsator with a modest amount of convective core overshooting.

Not long after the early studies by \citet{Degroote2010a} and \citet{Aerts2011}, another important result for massive star asteroseismology from the CoRoT mission was made based on the O9\,V star HD~46202 \citep{Briquet2011}. Spectroscopy of HD~46202 confirmed its literature spectral type and yielded a projected surface rotation rate of $v\,\sin\,i \simeq 25$~km\,s$^{-1}$ and an effective temperature of $T_{\rm eff} = 34\,100 \pm 600$~K, which makes it amongst the hottest $\beta$~Cep pulsators known \citep{Briquet2011}. Several radial and non-radial pulsation modes were identified using the CoRoT photometry of HD~46202. Forward seismic modelling of these identified modes yielded a precise mass and age for this massive pulsator, confirming it as the most massive modelled $\beta$~Cep star to date. A significant amount of convective core overshooting was required to fit the observed pulsation frequencies and the best-fitting asteroseismic parameters from \citet{Briquet2011} are provided in Table~\ref{table: modelling}.

A third pivotal study of a massive star using asteroseismology from CoRoT data is the magnetic and fast-rotating B3.5\,V star HD~43317 \citep{Papics2012a, Briquet2013, Buysschaert2017b, Buysschaert2018c}. This star was originally studied by \citet{Papics2012a} using the 150-d CoRoT light curve and it was concluded to exhibit g and p modes based on its frequency spectrum, since it appeared to have independent pulsation modes in both high and low frequency regimes. High-resolution spectroscopy confirmed HD~43317 as an early-type star with an effective temperature of $T_{\rm eff} = 17\,350 \pm 750$~K, and as a fast rotator with a projected surface rotational velocity of $v\,\sin\,i = 115 \pm 9$~km\,s$^{-1}$ \citep{Papics2012a}. Later, HD~43317 was detected to host a large-scale magnetic field with a polar field strength of approximately 1.3~kG \citep{Briquet2013, Buysschaert2017b}. After revisiting the light curve with the knowledge of how such fast rotation perturbs the frequencies of g-mode pulsations in B~stars (see e.g. \citealt{Kurtz2015b}), \citet{Buysschaert2018c} extracted all significant pulsation mode frequencies and performed forward seismic modelling of HD~43317 and determined its fundamental parameters, which are provided in Table~\ref{table: modelling}. The small amount of convective core overshooting in the rapidly-rotating and magnetic star HD~43317 led \citet{Buysschaert2018c} to a similar conclusion to that of \citet{Briquet2011} for the $\beta$~Cep star V2052~Oph (cf. \ref{subsection: ground}): the presence of a large-scale magnetic field is a plausible cause for the suppression of additional mixing (i.e. overshooting) in the near-core region.

As discussed in section~\ref{subsection: Kepler}, during the life time of the {\it Kepler} space mission, massive stars were avoided as targets in the field of view. Hence, whereas much of the asteroseismic inference of massive stars based on ground-based data were focussed on $\beta$~Cep stars, more recent asteroseismic results based on the 4-yr light curves of the {\it Kepler} telescope are typically derived from SPB stars \citep{Aerts2019b}. Although not all SPB stars are ``massive'' stars, they do share a common interior structure whilst on the main sequence, so their discussion is relevant as part of this review. 

Amongst the early days of {\it Kepler} asteroseismology, two SPB stars were identified as high priority targets: KIC~10526294 and KIC~7760680. A long series of rotationally-split g~modes in the B8.3\,V star KIC~10526294 allowed \citet{Papics2014} to determine a near-core rotation period of approximately 188~d. Later, \citet{Moravveji2015b} performed the first in-depth asteroseismic modelling of this main-sequence B star, from which it was concluded that a modest amount of exponential diffusive overshooting (i.e. $f_{\rm ov}$) fit the {\it Kepler} data significantly better than when using the step overshooting prescription (i.e. $\alpha_{\rm ov}$). Furthermore, a small but non-negligible amount of additional envelope mixing ($D_{\rm mix} \simeq 100$~cm$^2$\,s$^{-1}$) was needed despite KIC~10526294 being an ultra-slow rotator \citep{Moravveji2015b}. The best-fitting asteroseismic modelling parameters, including the measured overshooting of $0.017 \leq f_{\rm ov} \leq 0.018$, are included in Table~\ref{table: modelling}. The exquisite {\it Kepler} data allowed \citet{Triana2015} to compute a near-rigid interior rotation profile through an asteroseismic inversion. Interestingly, at the 1-$\sigma$ confidence level, the inversion by \citet{Triana2015} confirmed the ultra-slow rotation rate, but revealed that KIC~10526294 had a counter-rotating envelope compared to that of its core. Similar conclusions in terms of a non-negligible amount of convective core overshooting and envelope mixing were also reached for the B8~V SPB star KIC~7760680 based on the period spacing pattern of prograde dipole g~modes by \citet{Moravveji2016b}. However, it should be noted that the uncertainties obtained by \citet{Moravveji2015b} and \citet{Moravveji2016b} are typically quoted as the step size within the computed grid of evolution models and may be unrealistically small, because they ignore degeneracies amongst the model parameters (see \citealt{Aerts2018b}). There has been substantial development in the correct way to treat parameter correlations and degeneracies for asteroseismology of g-mode pulsators by moving away from the often-used $\chi^{2}$ merit function and towards the use of the Mahalanobis distance which includes heteroscedasticity. We refer the reader to \citet{Aerts2018b} for further details. 

At somewhat higher masses, \citet{Szewczuk2018a} performed forward seismic modelling of the SPB star KIC~3240411 using its axisymmetric ($m=0$) period spacing pattern extracted from {\it Kepler} data. KIC~3240411 was found to be a relatively young star near the zero-age main sequence with the upper limit on its convective core overshooting being $f_{\rm ov} \leq 0.030$. The best-fitting model parameters (i.e. model \#2 in Table~2 of \citealt{Szewczuk2018a}) are included in Table~\ref{table: modelling}. Importantly, \citet{Szewczuk2018a} also studied the effect of rotation and opacity data within their modelling for the purposes of explaining mode excitation of high-radial order g~modes in such stars. This is because the excitation of such a large number and radial order range for high-radial order g~modes remains somewhat difficult to explain from a theoretical perspective.

Recently, the nanosatelittes of the BRITE-constellation mission have been targeting pulsating massive stars with the aim of performing forward seismic modelling of the detected pulsation mode frequencies. Notable examples of the importance of the contribution of the BRITE mission in this aspect include $\nu$~Eri \citep{Daszy2017a} and $\theta$~Oph \citep{Walczak2019a}. The BRITE data marked the first time that these famous $\beta$~Cep stars were observed from space, thus the first time an asteroseismic analysis of them benefitted from having near-continuous, short-cadence and long-term monitoring. \citet{Handler2017a} discovered new g-mode pulsations in $\nu$~Eri using BRITE data, and \citet{Daszy2017a} used the extracted pulsation mode frequencies to perform forward seismic modelling. \citet{Daszy2017a} determined optimum parameters for $\nu$~Eri and emphasised the improvement of using modified opacity tables to explain all the observed pulsation mode frequencies. Moreover, the modified opacity data increases the efficiency of convection within the Z-bump, an effect also predicted when increasing the metallicity of the star \citep{Cantiello2009a} or increasing the rotation \citep{Maeder2008a}. In the analysis of $\theta$~Oph using BRITE data, \citet{Walczak2019a} discovered several g~modes. From their forward seismic modelling, similar conclusions made by \citet{Daszy2017a} for $\nu$~Eri were made. Specifically that an increase in the mean opacity within the Z-bump is needed to explain all pulsation mode frequencies in $\theta$~Oph \citep{Walczak2019a}. However, since $\theta$~Oph is a triple system with $\sim8.5$ and $\sim5.5$~M$_{\odot}$ primary and tertiary components, respectively, it was difficult to ascertain the origin of the g-mode frequencies \citep{Walczak2019a}.

Recently, a new method of performing forward seismic modelling by means of a machine learning was developed by \citet{Hendriks2019}. The authors trained a deep neural network using more than 62~million pulsation mode frequencies, which were calculated from a vast grid of stellar models covering main sequence intermediate- and high-mass stars spanning from 2 to 20~M$_{\odot}$. Optimum models were selected using a genetic algorithm making such an automated pipeline extremely quick compared to more conventional forward seismic modelling techniques. \citet{Hendriks2019} test their methodology using several well-studied massive stars to benchmark the accuracy of their technique and good agreement is found overall. However, as discussed in detail by \citet{Hendriks2019}, the modelling results based on their deep neural network depend on the choice of hyperparameters, which affect the ability to find the global minimum in the solution space. In Table~\ref{table: modelling}, the model parameters resulting from the optimised tuning of these hyperparameters are provided (cf. gray points), but such values are claimed to be an ``optimal starting point'' for further complex modelling. We refer the reader to \citet{Hendriks2019} for full details.

Asteroseismology using {\it Kepler} data has also been applied to hundreds of intermediate-mass stars covering masses between approximately 1 and 8~M$_{\odot}$, rotation rates up to 80\% of critical, and evolutionary stages spanning from the main sequence through to the red giant branch. More specifically, g-mode period spacings have also been used to probe interior rotation in hundreds of intermediate-mass main-sequence stars, which are more commonly known as $\gamma$~Dor stars \citep{VanReeth2015a, VanReeth2015b, VanReeth2016a, VanReeth2018a, Ouazzani2017a, Ouazzani2019a, Li_G_2019a, Li_G_2019b, Li_G_2020a}. An important conclusion from these works is that current angular momentum transport theory is erroneous by more than an order of magnitude \citep{Aerts2019b}. The situation is less clear for massive stars owing to the much smaller sample size currently available, but significant progress has already been made in recent years because of ground- and space-based data sets and asteroseismology. 

The pioneering asteroseismic studies of massive stars using ground- and space-based data sets clearly demonstrate the importance of constraining the interior properties of massive stars and the effectiveness of using g-mode period spacing patterns in SPB stars and low-radial order g and p modes for $\beta$~Cep stars. Thanks to the ongoing TESS mission \citep{Ricker2015}, the asteroseismic sample of pulsating massive stars has increased in size by at least two orders of magnitude \citep{Pedersen2019a, Burssens2020a}. Thus detailed forward seismic modelling of many high-mass TESS targets are expected in the not-so-distant future. We refer the reader to \citet{Handler2019a} for the first modelling study of a $\beta$~Cep star using TESS data. Despite the relatively small sample size so far compared to intermediate-mass stars, asteroseismic studies of massive stars have demonstrated the need to include improved prescriptions for convective core overshooting and envelope mixing given that current evolution models underestimate the core masses of massive stars and consequently also their main-sequence lifetimes.

\vspace{0.5cm}


\subsection{Implications of mixing for the post-main sequence}
\label{subsection: post-ms}

\begin{figure}[ht!]
\centering
\includegraphics[width=0.5\textwidth,clip]{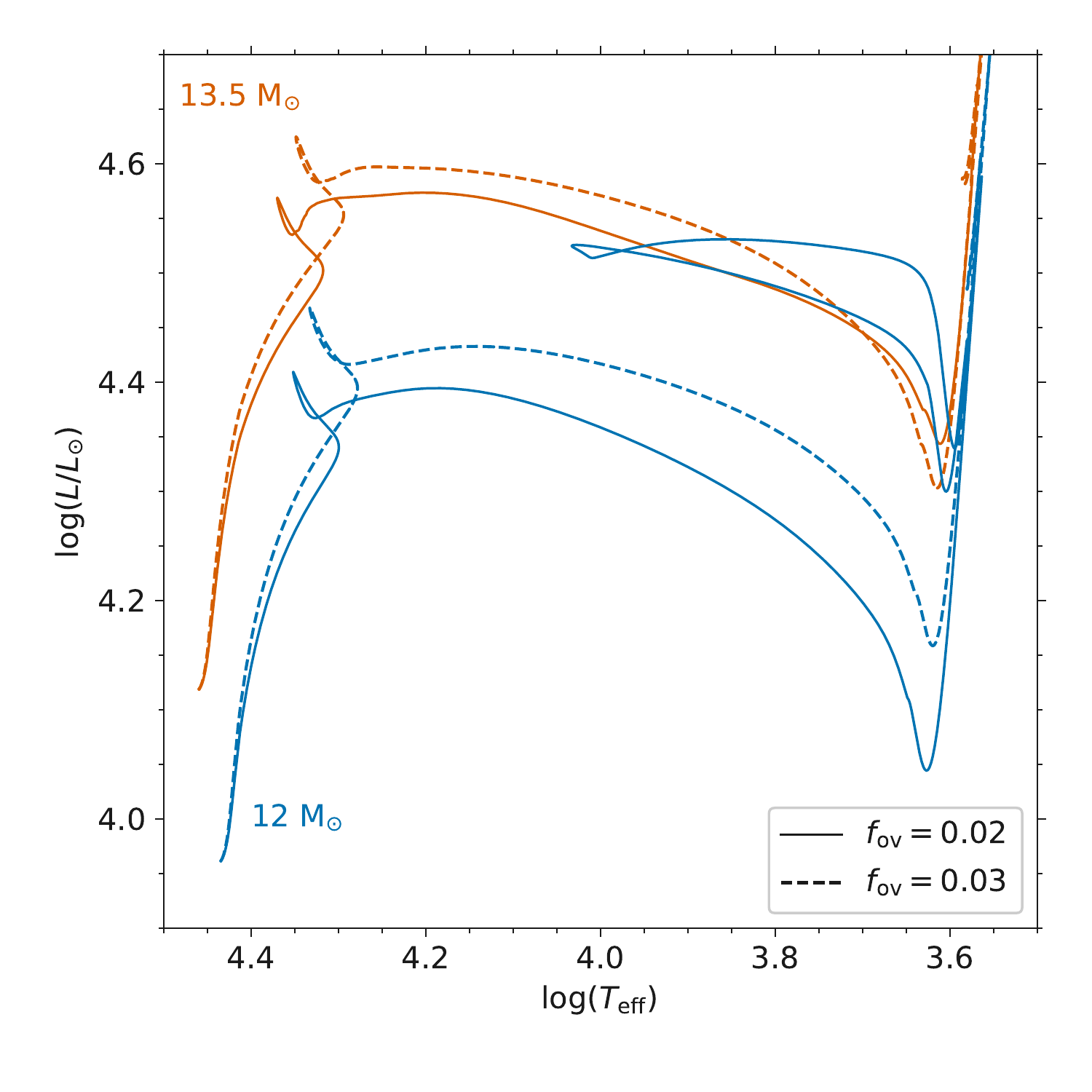}      
\caption{Hertzsprung-Russell diagram containing non-rotating evolutionary tracks starting from the zero-age main-sequence for stars of initial masses of 12- and 13.5-M$_{\odot}$ calculated using the MESA stellar evolution code \citep{Paxton2019}, using two different values for the diffusive exponential convective core-overshooting value ($f_{\rm ov}$), which are expressed in pressure scale heights. With increased mixing in the near-core region, a massive star has a longer main-sequence lifetime and produces a larger convective core mass at the terminal age main-sequence. The increased core mass and interior mixing during the main sequence phase of evolution also impacts the post-main sequence evolution of a massive star, including whether it performs a {\it ``blue loop''}, which is demonstrated by the evolutionary track of the 12-M$_{\odot}$ star with $f_{\rm ov} = 0.02$.}
\label{figure: MESA}
\end{figure}

The majority of asteroseismic studies of massive stars have been of main sequence stars. Of course, based on evolutionary time scales, stars spend more than 90\% of their lives in this phase \citep{Kippenhahn_BOOK}. However, pulsating post-main sequence massive stars do exist \citep{Saio2006b, ASTERO_BOOK}, and to truly understand their interior physics requires us to first understand the interiors of main-sequence stars. For example, the interplay of different mixing processes can cause {\it some} post-main sequence massive stars to undergo blue loops in the HR~diagram. However, the exact nature of why some massive stars undergo blue loops and others do not remains unknown \citep{Langer2012}, especially given the differences in the physics and numerics in various stellar evolution codes (e.g. \citealt{Martins2013c}). It is known, however, that convection, mixing and mass loss all play significant roles in determining if stars undergo blue loops in the HR~diagram and their pulsational properties \citep{Georgy2012a, Georgy2014a, Saio2013c, Wagle2019a}. To demonstrate how the difference in the amount of the mixing at the boundary of a convective core within a massive main-sequence star drastically impacts its post-main sequence evolution, evolutionary tracks calculated with the MESA stellar evolution code (\citealt{Paxton2011, Paxton2013, Paxton2015, Paxton2018, Paxton2019}; r11554) for initial masses of 12- and 13.5-M$_{\odot}$ and two moderate values for their convective core-overshooting value using the diffusive exponential prescription (i.e. $f_{\rm ov}$) are shown in Fig.~\ref{figure: MESA}. The evolutionary tracks shown in Fig.~\ref{figure: MESA} use initial hydrogen and metal mass fractions of X = 0.71 and Z = 0.02, respectively, OP opacity tables \citep{Seaton2005}, and the chemical mixture of \citet{Nieva2012} and \citet{Przybilla2013b} for cosmic B stars. As illustrated in Fig.~\ref{figure: MESA}, only one of the four evolution tracks contains a blue loop in the post-main sequence stage of evolution.

Currently, there exists a knowledge gap concerning the interiors of blue supergiants yet to be filled by asteroseismology. This is primarily because until recently few pulsating blue supergiants had been found within the Hertzsprung gap in the HR~diagram \citep{Bowman2019b}, despite theoretical models predicting such stars should pulsate \citep{Saio2013c, Daszy2013c, Ostrowski2015b}. Asteroseismology of post-main sequence massive stars represents a major future goal for the community, especially since g-mode period spacing patterns and the presence of radial p~modes have the capability to distinguish shell-hydrogen burning and core-helium burning massive stars \citep{Saio2013c, Bowman2019b}. Furthermore, when coupled to spectroscopic surface abundances of core-processed material (e.g. C, N and O), convection and mixing in massive stars can be significantly improved within stellar evolution codes, which has important consequences for stars that eventually explode as a supernova (e.g. \citealt{Saio1988b, Georgy2014a, Wagle2019a}). In particular, the helium core masses of post-main sequence massive stars represent a critical deliverable to the wider astronomical community as it is a fundamental parameter for predicting the chemical properties of supernovae \citep{Smartt2009b, Langer2012}.


\section{Diverse photometric variability in massive stars}
\label{section: waves}

\begin{figure}[ht!]
\centering
\includegraphics[width=0.5\textwidth,clip]{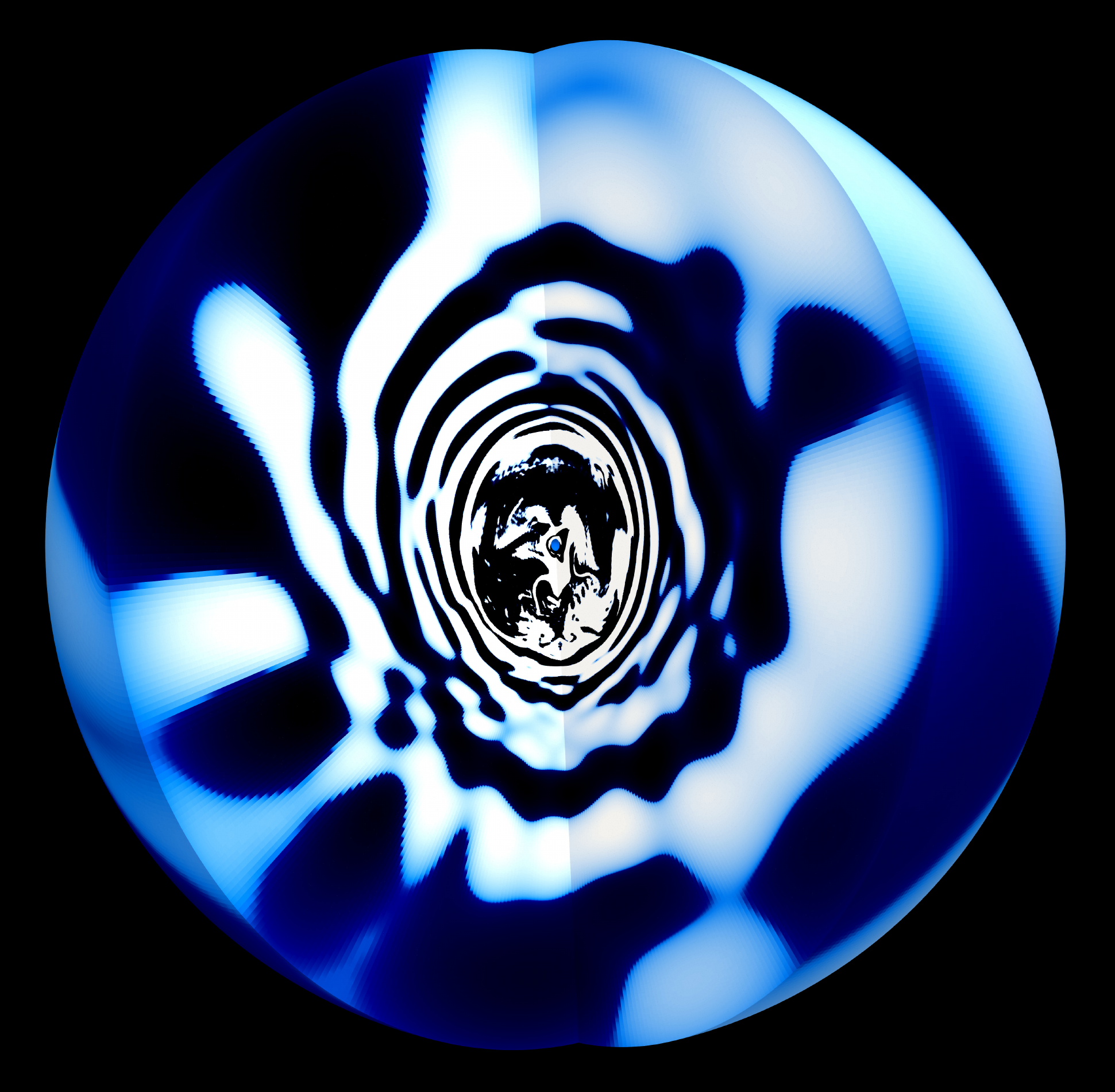}      
\caption{Snapshot of a non-rotating 3D numerical simulation of gravity waves excited by core convection inside a main-sequence massive star, with a white-blue colour scale indicating temperature fluctuations. Simulation courtesy of \citet{Edelmann2019a}.}
\label{figure: 3D simulation}
\end{figure}

A recent discovery made using the CoRoT, K2 and TESS missions was that the vast majority of early-type stars have significant low-frequency variability in photometry \citep{Bowman2019a, Bowman2019b}. Such stochastic variability is not predicted from the $\kappa$-mechanism, but is expected from convectively-driven internal gravity waves (IGWs) excited by core convection \citep{Rogers2013b, Rogers2015, Edelmann2019a, Horst2020a*}. 2D and 3D hydrodynamical simulations predict that IGWs reach the surface with significant amplitudes and provide detectable perturbations in temperature and velocity \citep{Edelmann2019a}, and that IGWs are also extremely efficient at transporting angular momentum and chemical elements \citep{Rogers2015, Rogers2017c}. A snapshot of a 3D simulation of IGWs propagating within a 3-M$_{\odot}$ main-sequence star from \citet{Edelmann2019a} is shown in Figure~\ref{figure: 3D simulation}.

The morphology of the low-frequency variability in more than 160 massive stars was found to be similar across a large range of masses and ages for both metal-rich galactic and metal-poor LMC stars \citep{Bowman2019b}. The insensitivity of the stochastic variability to a star's metallicity together with the fact that evolutionary timescales predicted most stars were likely on the main sequence was concluded as strong evidence that the observed stochastic variability is likely caused by IGWs excited by core convection \citep{Bowman2019b}. More recently, \citet{Bowman2020b} demonstrated that the morphology of the IGWs in the frequency spectrum probes the evolutionary properties of the star, such as mass and radius. Furthermore, the amplitudes of IGWs in photometry were found to correlate with the macroturbulent broadening in the spectral lines of dozens of massive stars observed by the TESS mission \citep{Bowman2020b}, with macroturbulence also having been associated with pulsations \citep{Aerts2009b, Simon-Diaz2010b, Simon-Diaz2017a}. Thus, mixing and angular momentum transport caused by IGWs are important to take into account when studying massive star evolution \citep{Aerts2019b}, and are currently not implemented in most stellar evolution codes.

There are currently four known excitation mechanisms that can trigger variability in early-type stars: (i) coherent p and g~modes excited by the $\kappa$-mechanism \citep{Dziembowski1993f, Gautschy1993a, Szewczuk2017a}; (ii) IGWs generated by turbulent core convection \citep{Edelmann2019a}; (iii) IGWs generated by thin sub-surface convection zones \citep{Cantiello2009a}; and (iv) waves generated from tides in binary systems \citep{Fuller2017c}. Whereas essentially all of the asteroseismic results discussed in section~\ref{section: constraints} are based on heat-driven modes, the other variability mechanisms in massive stars remain somewhat under-utilised despite having great probing power of stellar interiors. In the future, it is expected that binary asteroseismology in particular is going to play a more substantial role in providing excellent constraints of massive star interiors because recent space telescopes such as BRITE and TESS are providing the necessary high-quality time-series observations (see e.g. \citealt{Jerzykiewicz2020a} and \citealt{Southworth2020a}).


\section{Conclusions and Future Prospects}
\label{section: conclusions}

Our knowledge of the interior physics of massive stars has historically been based on a handful of asteroseismic studies using time series data obtained with ground-based telescopes. Such data sets are notoriously limited by their poor duty cycles which complicates mode identification \citep{Aerts2003d, Dupret2004b, Handler2004b, Pamyat2004, Ausseloos2004, Briquet2007e, Briquet2012, Desmet2009b, Daszy2013b, Daszy2017a}. These initial asteroseismic studies, however, provided the first evidence of the interior rotation profiles of massive stars and that evolutionary models of such stars lacked calibrated prescriptions for interior processes such as convective core overshooting. Consequently standard evolutionary models typically underestimate the mass of convective cores and hence the main sequence lifetimes of massive stars. Furthermore, ground-based studies yielded the first measurements of the interior rotation rates of main sequence stars \citep{Aerts2003d}, with both rigid and non-rigid rotation rates being inferred for a handful of $\beta$~Cep stars \citep{Aerts2003d, Pamyat2004, Ausseloos2004, Briquet2007e}. More recently, the MOST \citep{Walker2003}, CoRoT \citep{Auvergne2009}, Kepler/K2 \citep{Borucki2010, Howell2014}, and the ongoing BRITE-Constellation \citep{Weiss2014} missions have advanced asteroseismic studies beyond measuring only the masses, ages and metallicities of massive stars. More advanced forward seismic modelling techniques and improved observations have constrained both macroscopic and microscopic physical processes and mechanisms inside stars, such as the amount and shape of mixing in their radiative envelopes \citep{Moravveji2015b, Moravveji2016b, Buysschaert2018c, Szewczuk2018a, Walczak2019a} and angular momentum transport as a function of stellar evolution \citep{Aerts2019b}.

Today, thanks to the ongoing TESS mission \citep{Ricker2015}, there is huge asteroseismic potential for massive stars as we are no longer limited by the biases of having a small sample of pulsating massive stars. The long-term and high-photometric precision provided by space telescopes is unrivalled by ground-based telescopes, and the sample of massive stars has expanded to hundreds of stars because of the all-sky TESS mission \citep{Pedersen2019a, Burssens2020a}. Crucially, TESS is observing massive stars which span a large range in mass and age, but also massive stars in different metallicity regimes. This is because the southern CVZ of TESS includes the LMC galaxy, which allows pulsation excitation models to be tested for metal-rich and metal-poor stars \citep{Bowman2019b}. TESS also offers the opportunity to revisit ``old friends'' in terms of previously studied massive stars with high-precision photometry, which is particularly useful to probe the long-term stability in pulsation mode amplitudes and frequencies in relatively short-lived stars (e.g. \citealt{Neilson2015b}). The diverse variability of massive stars, which includes both coherent pulsation modes excited by the $\kappa$-mechanism and IGWs excited by core convection \citep{Pedersen2019a, Bowman2019b, Burssens2020a}, enables asteroseismology for a sample of massive stars larger by two orders of magnitude compared to any that came before.

The important future goals of asteroseismic studies based on the large and diverse TESS data set include constraining the helium core masses, near-core and envelope mixing processes, interior rotation profiles and angular momentum transport mechanisms inside massive stars. Insight of the physics in the near-core region of stars above approximately 8~M$_{\odot}$ is currently lacking compared to intermediate- and low-mass stars \citep{Aerts2019b}. Moreover, HD~46202 remains the only massive star above 15~M$_{\odot}$ to have undergone forward seismic modelling \citep{Briquet2011}. In turn TESS data combined with high-resolution spectroscopy and asteroseismology will mitigate the currently large uncertainties in stellar evolution theory and lead to constraining why only some massive stars undergo blue loops during the post-main sequence phase of evolution (e.g. \citealt{Bowman2019b}), and improved predictions of supernovae chemical yields and remnant masses. The future is very bright for massive stars, and the goal to calibrate stellar structure and evolution models of massive stars using asteroseismology is now within reach.


\section*{Conflict of Interest Statement}
The author declares that the research was conducted in the absence of any commercial or financial relationships that could be construed as a potential conflict of interest.


\section*{Author Contributions}
The author confirms being the sole contributor of this work and has approved it for publication.


\section*{Funding}
DMB gratefully acknowledges funding from the European Research Council (ERC) under the European Union's Horizon 2020 research and innovation programme (grant agreement No. 670519: MAMSIE), and a senior postdoctoral fellowship from the Research Foundation Flanders (FWO) with grant agreement No.~1286521N.


\section*{Acknowledgments}
DMB is grateful to S.~Burssens who graciously re-produced Fig.~\ref{figure: TESS HRD} and to P.~V.~F.~Edelmann who made Fig.~\ref{figure: 3D simulation}. DMB is grateful to the anonymous reviewers and Prof. C. Aerts for constructive feedback. DMB is also grateful to the {\sc mesa} and {\sc gyre} developers, in particular B.~Paxton and R.~H.~D.~Townsend, for continually supporting the development of state-of-the-art and open-source tools for modelling massive stars. The Kepler, K2 and TESS data presented in this paper were obtained from the Mikulski Archive for Space Telescopes (MAST) at the Space Telescope Science Institute (STScI), which is operated by the Association of Universities for Research in Astronomy, Inc., under NASA contract NAS5-26555. Support to MAST for these data is provided by the NASA Office of Space Science via grant NAG5-7584 and by other grants and contracts. Funding for the Kepler/K2 mission was provided by NASA’s Science Mission Directorate. Funding for the TESS mission is provided by the NASA Explorer Program. Funding for the TESS Asteroseismic Science Operations Centre is provided by the Danish National Research Foundation (Grant agreement no.: DNRF106), ESA PRODEX (PEA 4000119301) and Stellar Astrophysics Centre (SAC) at Aarhus University, Denmark. This research has made use of the SIMBAD database and the VizieR catalog access tool operated at CDS, Strasbourg, France, and the SAO/NASA Astrophysics Data System. The research leading to these results has received funding from the European Research Council (ERC) under the European Union's Horizon 2020 research and innovation programme (grant agreement No.~670519: MAMSIE), and a senior postdoctoral fellowship from the Research Foundation Flanders (FWO) with grant agreement No.~1286521N.


\newpage
\begin{landscape}
\begin{table*}[ht!]
\centering
\tiny
\caption[]{Best-fitting parameters of high-mass stars derived from forward seismic modelling of pulsations discussed in this review. Note that not all studies listed here use the same evolution codes, opacity tables (e.g. standard versus modified), nor necessarily the same numerical setup for selecting the statistically best-fitting asteroseismic model, so we refer the reader to each individual study for full details.}
\begin{tabular}{l c c c c c c r}
\hline
\multicolumn{1}{c}{Star} & \multicolumn{1}{c}{Mass} & \multicolumn{1}{c}{Age} & \multicolumn{1}{c}{$Z$} & \multicolumn{1}{c}{$X_{c}$} & \multicolumn{2}{c}{Core overshooting} & \multicolumn{1}{c}{Reference} \\

\multicolumn{1}{c}{} & \multicolumn{1}{c}{} & \multicolumn{1}{c}{} & \multicolumn{1}{c}{} & \multicolumn{1}{c}{} & \multicolumn{1}{c}{$\alpha_{\rm ov}$} & \multicolumn{1}{c}{$f_{\rm ov}$} & \multicolumn{1}{c}{} \\

\multicolumn{1}{c}{} & \multicolumn{1}{c}{(M$_{\odot}$)} & \multicolumn{1}{c}{(Myr)} & \multicolumn{1}{c}{(dex)} & \multicolumn{1}{c}{(dex)} & \multicolumn{2}{c}{} & \multicolumn{1}{c}{} \\

\hline 

V836~Cen (HD~129929)			&	$9.35$	&	$16.3$	&	$0.0188$	&	$0.353$		&	$0.10 \pm 0.05$	&	$-$	&	\citet{Dupret2004b}	\\
							&	$9.23$	&	$-$	&	$0.0120$	&	$0.314$	&	$-$	&	$0.0365$	&	\citet{Hendriks2019}	 \\
\hline
$\nu$~Eri (HD~29248)		&	$9.0-9.9$	&	$16-20$	&	$0.0150$	&	$0.34-0.38$	&	$0.00-0.12$		&	$-$	&	\citet{Pamyat2004}	\\
							&	$8.4$	&	$26.6$	&	$0.0115$	&	$-$			&	$0.21$			&	$-$	&	\citet{Ausseloos2004}	\\
							&	$7.13$	&	$14.82$	&	$0.019$	&	$0.139$	&	$0.28$		&	$-$	&	\citet{Suarez2009a}	\\
							&	$8.01-9.77$	&	$16.2-29.1$	&	$0.014-0.018$	&	$0.247-0.279$	&	$0.03-0.35$		&	$-$	&	\citet{Daszy2010a}	\\			
							&	$9.0$	&	$-$	&	$0.015$	&	$-$	&	$0.163$		&	$-$	&	\citet{Daszy2017a}	\\
							&	$6.56$	&	$-$	&	$0.0195$	&	$0.275$	&	$-$	&	$0.0242$	&	\citet{Hendriks2019}	 \\
\hline
$\theta$~Oph (HD~157056)		&	$8.2 \pm 0.3$	&	$-$	&	$0.009-0.015$	&	$0.38 \pm 0.02$	&	$0.44 \pm 0.07$	&	$-$	&	\citet{Briquet2007e}	\\
							&	$8.00-8.81$	&	$7.20-7.38$	&	$0.01-0.02$	&	$0.35-0.42$	&	$0.07-0.39$	&	$-$	&	\citet{Walczak2019a}	 \\
							&	$7.21$	&	$-$	&	$0.0132$	&	$0.349$	&	$-$	&	$0.0402$	&	\citet{Hendriks2019}	 \\
\hline
V2052~Oph (HD~163472)			&	$8.2-9.6$	&	$16.9-23.7$	&	$0.010-0.016$	&	$0.25-0.32$	&	$0.00-0.15$	&	$-$	&	\citet{Briquet2012}	\\
\hline
12~Lac (HD~214993)			&	$10.2-14.4$	&	$11-23$	&	$0.015$	&	$0.13-0.21$		&	$0.0-0.4$	&	$-$	&	\citet{Desmet2009b}	\\
							&	$10.28$	&	$-$	&	$0.0115$	&	$-$		&	$0.39$	&	$-$	&	\citet{Daszy2013b}	\\
							&	$12.80$	&	$-$	&	$0.0198$	&	$0.200$	&	$-$	&	$0.0187$	&	\citet{Hendriks2019}	 \\
\hline
HD~50230					&	$7-8$	&	$-$	&	$0.020$	&	$\simeq0.30$	&	$0.2-0.3$	&	$-$	&	\citet{Degroote2010a}	\\
							&	$11.12$	&	$-$	&	$0.0200$	&	$0.045$	&	$-$	&	$0.0494$	&	\citet{Hendriks2019}	 \\
							&	$6.187 \pm 0.025$	&	$61.72^{+1.89}_{-0.21}$	&	$0.0408 \pm 0.0009$	&	$0.3058^{+0.0006}_{-0.0007}$	&	$-$	&	$0.0180 \pm 0.0014$	&	\citet{Wu_T_2019a}	\\

\hline
HD~180642					&	$11.4-11.8$	&	$12.4-13.0$	&	$0.008-0.014$	&	$0.21-0.25$	&	$<0.05$	&	$-$	&	\citet{Aerts2011} \\
\hline
HD~46202					&	$24.1 \pm 0.8$	&	$4.3 \pm 0.5$	&	$0.013-0.015$	&	$-$	&	$0.10 \pm 0.05$	&	$-$	&	\citet{Briquet2011}	\\
\hline
KIC~10526294					&	$3.25$	&	$63$	&	$0.014$	&	$0.627$	&	$-$	&	$0.017$	&	\citet{Moravveji2015b} \\
							&	$5.25$	&	$-$	&	$0.0120$	&	$0.712$	&	$-$	&	$0.0348$	&	\citet{Hendriks2019}	 \\
\hline
KIC~	7760680					&	$3.25 \pm 0.05$	&	$202$	&	$0.020 \pm 0.001$	&	$0.503 \pm 0.001$	&	$-$	&	$0.024 \pm 0.001$	&	\citet{Moravveji2016b} \\
\hline
HD~43317					&	$5.8^{+0.2}_{-0.1}$	&	$$	&	$$	&	$0.54^{+0.01}_{-0.02}$	&	$-$	&	$0.004^{+0.014}_{-0.002}$	&	\citet{Buysschaert2018c} \\
\hline
KIC~3240411					&	$6.25$	&	$-$	&	$0.006$	&	$0.612$	&	$-$	&	$0.02$	&	\citet{Szewczuk2018a} \\
\hline
\end{tabular}
\label{table: modelling}
\end{table*}
\end{landscape}
\newpage


\newcommand{\apss}{\it Astrophys. Space Sci.}
\newcommand{\aaps}{\it Astron. Astrophys. Sup. Ser.}
\newcommand{\actaa}{\it Astron. Nach.}
\newcommand{\aapr}{\it Astron. Astrophys. Rev.}
\newcommand{\araa}{\it Ann. Rev. Astron. Astrophys.}
\newcommand{\aap}{\it Astron. Astrophys.}
\newcommand{\aj}{\it Astrom. J. }
\newcommand{\apj}{\it Astrophys. J.}
\newcommand{\apjl}{\it Astrophys. J. Letters}
\newcommand{\apjs}{\it Astrophys. J. Supp. Ser.}
\newcommand{\mnras}{\it Mon. Not. Roy. Astron. Soc.}
\newcommand{\pasj}{\it Public. Astronom. Soc. Jpn.}
\newcommand{\pasp}{\it Public. Astronom. Soc. Pac.}
\newcommand{\ssr}{Space Science Review}
\newcommand{\prx}{\it Phys. Rev. X}
\newcommand{\prl}{\it Phys. Rev. Letters}
\newcommand{\nat}{\it Nature}

\bibliographystyle{frontiersinSCNS_ENG_HUMS} 
\bibliography{Bowman_frontiers_accepted_arxiv}


\end{document}